\documentclass[10pt,journal,compsoc]{IEEEtran}
\usepackage{sushovan}
\usepackage{comment}
\usepackage{url}
\usepackage[nocompress]{cite}
\usepackage{tabularx}
\ifCLASSOPTIONcompsoc
  \usepackage[nocompress]{cite}
\else
  \usepackage{cite}
\fi
\ifCLASSINFOpdf
\else
\fi
\begin{document}
\def\mvp{\vspace*{-0.1in}}
\title{BayesWipe: A Scalable Probabilistic Framework for Cleaning BigData}
\author{
      Sushovan~De,
      Yuheng~Hu,
      Meduri~Venkata~Vamsikrishna,
      Yi~Chen,
      and~Subbarao~Kambhampati%
\IEEEcompsocitemizethanks{
\IEEEcompsocthanksitem{This work was done when all the authors were
  with the Department of Computer Science \& Engineering at Arizona
  State University, Tempe, AZ 85287. Sushovan De is now with Google
  Inc. Yuheng Hu is now with IBM Research, Almaden. Yi Chen is now
  with the School of Management and the College of Computing at New
  Jersey Institute of Technology.}}}
%
\IEEEtitleabstractindextext{%
\begin{abstract}
Recent efforts in data cleaning of structured data have focused exclusively on
problems like data deduplication, record matching, and data standardization;
none of the approaches addressing these problems focus on fixing incorrect attribute values in tuples. Correcting
values in tuples is typically performed by a minimum cost repair of tuples
that violate static constraints like CFDs (which have to be provided by domain
experts, or learned from a clean sample of the database). In this paper, we
provide a method for correcting individual attribute values in a structured
database using a Bayesian generative model and a statistical error model
learned from the noisy database directly. We thus avoid the necessity for a
domain expert or clean master data. We also show how to efficiently perform
consistent query answering using this model over a dirty database, in case
write permissions to the database are unavailable. We evaluate our methods
over both synthetic and real data.
\end{abstract}
}

\maketitle
\section{Introduction}
\label{sec:introduction}


\IEEEPARstart{A}{lthough} data cleaning has been a long standing problem, it has become
critically important again because of the increased interest in big data and
web data.
Most of the focus of the work on big data has been on the volume, velocity, or
variety  of the data; however, an important part of making big data useful is
to ensure the veracity of the data. Enterprise data is known to have a typical
error rate of 1--5\%  \cite{fan2012foundations} (error rates of up to 30\%
have been observed). This has led to renewed interest in  cleaning of big data
sources, where manual data cleansing tasks are seen as prohibitively expensive
and time-consuming \cite{patrick2013before}, or the data has been generated by
users and cannot be implicitly trusted \cite{heather2010health}.
Among the various types of big data, the need to efficiently handle
large scaled structured data that is rife with inconsistency and
incompleteness is also more significant than ever. Indeed, multiple studies,
such as \cite{cra-bigdata} emphasize the importance of effective, efficient
methods for handling ``dirty big data''.

\begin{table}[!h]
\caption[A Snapshot of Car Data Extracted From the Web.]{A snapshot of car data extracted from \textsf{cars.com} using information extraction techniques}
\mvp

\begin{center}
\begin{tabular}{ | c || c | c | c | c | c | c | }
  \hline
  TID & Model & Make  & Orig & Size & Engine &  Condition\\
  \hline
  $t_1$ & \textsf{Civic}  & \textsf{Honda}    & \textsf{JPN} & \textsf{Mid-size} & \textsf{V4} & \textsf{NEW} \\
  $t_2$ & \textsf{Focus}  & \textsf{Ford}    & \textsf{USA}  & \textsf{Compact} & \textsf{V4} & \textsf{USED} \\
  $t_3$ & \textsf{Civik}    & \textsf{Honda}    & \textsf{JPN} & \textsf{Mid-size} & \textsf{V4} & \textsf{USED} \\
  $t_4$ & \textsf{Civic}    & \textsf{Ford}     & \textsf{USA} & \textsf{Compact}  & \textsf{V4} & \textsf{USED} \\
  $t_5$ &                   & \textsf{Honda}     & \textsf{JPN} & \textsf{Mid-size}  & \textsf{V4} & \textsf{NEW}  \\
  $t_6$ & \textsf{Accord}   & \textsf{Honda}     & \textsf{JPN} & \textsf{Full-size}  & \textsf{V6} & \textsf{NEW} \\
  \hline
\end{tabular}\label{tbl:exp1}
\end{center}
\end{table}
Most of the current techniques are based on deterministic rules, which have a number of problems:
Suppose that the user is interested in finding `Civic' cars from Table~\ref{tbl:exp1}.
Traditional data retrieval systems would return tuples $t_1$ and $t_4$
for the query, because they are the only ones that are a match for the
query term. Thus, they completely miss the fact that $t_4$ is in fact
a dirty tuple --- A Ford Focus car mislabeled as a Civic.
Additionally, tuple $t_3$ and $t_5$ would not be returned as a result tuples
since they have a typos or missing values, although they represent desirable
results. The objective of this work is to provide the true result set ($t_1, t_3, t_5$)
to the user. 

Although this problem has received significant attention over the years in the
traditional database literature, the state-of-the-art approaches fall far
short of an effective solution for big data and web data.
Traditional methods include outlier detection
\cite{knorr2000distance}, noise removal \cite{xiong2006enhancing},
entity resolution \cite{singla2006entity,xiong2006enhancing}, and
imputation \cite{fellegi1976systematic}.
Although these methods are
efficient in their own scenarios, their dependence on clean master data is
a significant drawback.

Specifically, state of the art approaches (e.g.,
\cite{bohannon2005cost,fan2009discovering,leo2011cleaning}) attempt to clean
data by exploiting patterns in the data, which they express in the form of
conditional functional dependencies (or CFDs). In the motivating example,
the fact that \textsf{Honda} cars have \textsf{`JPN'} as the origin of the manufacturer
would be an example of such a pattern.
However, these approaches depend on the availability of a clean data corpus
or an external reference table to
learn data quality rules or patterns before fixing the errors in the
dirty data. Systems such as ConQuer \cite{renee2005conquer} depend
upon a set of clean constraints provided by the user.
Such clean corpora or constraints may be easy to establish in a tightly
controlled enterprise environment but are infeasible for web data and big data.
One may attempt to learn data quality rules directly from the noisy data.
Unfortunately however, our experimental evaluation
shows that  even small amounts of noise severely impairs the ability to
learn useful constraints from the data.

To avoid dependence on clean master data, we propose a novel system
called \bw\ \cite{bwconf} that assumes that a statistical
process underlies the generation of clean data (which we call the \emph{data source
model}) as well as the corruption of data (which we call the \emph{data error
model}). The noisy data itself
is used to learn the parameters of these the generative and error
models, eliminating dependence on clean master data. Then, by
treating the clean value as a latent random variable,
\bw\ leverages these two learned models and automatically
infers its value through a Bayesian estimation.

We designed \bw\ so that it can be used in two different modes: a traditional
\emph{offline cleaning} mode, and a novel \emph{online query processing} mode.
The offline cleaning mode of \bw\ follows the classical data cleaning
model, where the entire database is accessible and can be cleaned {\em in situ}.
This mode is particularly useful when one has complete control over the data, 
and a one-time cleaning of the data is needed. Data warehousing scenarios such as data crawled from the web,
or aggregated from various noisy sources can be effectively cleaned in this mode.
The cleaned data can be stored either in a deterministic database,
or in a probabilistic database. If a probabilistic database is chosen
as the output mode, \bw\ stores not only the clean
version of the tuple it believes to be most likely correct one,
but the entire distribution over possible clean tuples. The choice of a probabilistic output mode for the cleaned tuples is
most useful for those scenarios where recall is very important for
further data processing on the cleaned tuples.

One of the features of the
offline mode of \bw\ is that a probabilistic database (PDB) can be generated as
a result of the data cleaning. In the first instance, notice that \bw\ was built
for deterministic databases. It can operate on a deterministic database and
produce a probabilistic cleaned database as an output. Probabilistic databases are complex and
unintuitive, because each single input tuple is mapped into a distribution
over resulting clean alternatives. We show how the top-$k$ results can be retrieved from a PDB while displaying 
the clean data that is comprehensible to the user.

The online query processing mode of \bw\ is motivated by
web data scenarios where it is impractical to create a local copy of the data
and clean it offline, either due to large size, high frequency of change,
or access restrictions. In such cases, the best way to obtain clean answers
is to clean the resultset as we retrieve it, which also provides us the opportunity
of improving the efficiency of
the system, since we can now ignore entire portions of the database
which are likely to be unclean or irrelevant to the top-$k$.
\bw\ uses a \emph{query rewriting system} that enables
it to efficiently retrieve only
those tuples that are important to the top-$k$ result set.
This rewriting approach is inspired by, and is a significant extension of
our earlier work on QPIAD system for handling data incompleteness
\cite{qpiad2009}. In big data scenarios, clean master data is rarely available, and write access is either unavailable, or undesirable due to the
efficiency and indexing concerns. The online mode is particularly suited to get clean results in such results. 

We implement \bw\ in a Map-Reduce architecture, so that we can run it very quickly for massive datasets. 
The architecture for parallelizing \bw\ is explained more fully in Sec~\ref{sec:mapreduce}. In short, there is a two-stage map-reduce architecture, where in the first stage, the dirty 
tuples are routed to a set of reducer nodes which hold the relevant candidate clean tuples for them. In the second stage, the resulting candidate clean tuples along with their 
scores are collated, and the best replacement tuple is selected from them.

To summarize our contributions, we:

\begin{itemize}
  \item Propose that data cleaning should be done using a principled, probabilistic approach.
  \item Develop a novel algorithm following those principles, which uses a Bayes network as the generative model and maximum entropy as the error model of the data.
  \item Develop novel query rewriting techniques so that this algorithm can also be used in a big data scenario.
  \item Develop a parallelized version of this algorithm using map-reduce framework.
  \item Empirically evaluate the performance of our algorithm using both controlled and real datasets.
\end{itemize}

The rest of the paper is organized as follows.
We begin by discussing the related work and then
describe the architecture of \bw\ in the next section, where
we also present the overall algorithm.
Section~\ref{sec:approach} describes the learning phase of \bw, where
we find the generative and error models. Section~\ref{sec:offline} describes the
offline cleaning mode, and the next section details the query
rewriting and online data processing.
We describe the parallelized version of \bw\ in Section~\ref{sec:mapreduce} and the results of the our empirical evaluation in Section~\ref{sec:experiments},
and then conclude by summarizing our contributions. Further details about \bw\ can be found in the thesis \cite{de2014unsupervised}.

\vspace{-0.07in}
\section{Related Work}
\label{sec:related}

Much of the work in data cleaning focused on deterministic
dependency relations such as FD, CFD, and INDs.  Bohannon \etal\ proposed using Conditional
Functional Dependencies (CFD) to clean data \cite{bohannon2007conditional,fan2008conditional}.
Indeed, CFDs are very effective in cleaning data. However, the precision and recall of
cleaning data with CFDs completely depends on the quality of the set of dependencies used for the
cleaning. As our experiments show, learning CFDs from dirty data produces very unsatisfactory
results. In order for CFD-based methods to perform well, they need to be learned from
a clean sample of the database \cite{fan2009discovering}  
which must be large enough to be representative of all the patterns
in the data. Finding such a large corpus of clean master data is a non-trivial problem, and is
infeasible in all but the most controlled of environments (like a corporation with high
quality data).

A recent variant on the deterministic dependency based cleaning by J.Wang \etal\ \cite{fixingrules} proposes using fixing rules which contain negative(possible 
errors) and positive(clean replacements) patterns for an attribute. However, there can be several ways in which a tuple can go wrong and the detection of the positive pattern requires clean master data. 
\bw\ on the other hand uses an error model to detect errors automatically and clean them in the absence of clean master data.
Recent work by J.Wang \etal\ \cite{Sampleclean} plugs in one of the rule based cleaning techniques to clean a sample of the entire data 
and use it as a guideline to clean the entire data. It is important to note that this method only caters to aggregate numerical queries whereas the 
online mode of \bw\ supports all types of SQL queries (not just aggregates) and returns clean result tuples.

Although it is possible to ameliorate some of the difficulties of
CFD/AFD methods by considering approximate versions of them, the work
in the uncertainty in AI  community demonstrated the semantic pitfalls
of handling uncertainty in this way. In particular, approximate
versions of CFDs/AFDs considered in works such as
\cite{golab-near-optimal-tableaux,cormode2009estimating} are similar to the
certainty factors approaches for handling uncertainty that were
popular in the heyday of expert systems, but whose semantic
inconsistencies are by now well-established (see, for example, Section 14.7.1 of \cite{russell2010artificial}). Because of this, in this paper we
focus on a more systematic probabilistic approach.

Even if a curated set of integrity constraints are provided, existing methods do not use
a probabilistically principled method of choosing a candidate correction.
They resort to either heuristic based methods, finding an approximate algorithm for the
least-cost repair of the database \cite{arenas1999consistent, 
bohannon2005cost, cong2007improving}; using a human-guided repair \cite{yakout2011guided}, or
sampling from a space of possible repairs \cite{beskales2013sampling}.
There has been work that attempts to guarantee a correct repair
of the database \cite{fan2010towards}, but they can only provide guarantees for corrections
of those tuples that are supported by data from a perfectly clean master database.
Recently, \cite{beskales2013relative} have shown how the relative trust one
places on the constraints and the data itself plays into the choice of cleaning tuples.
A Bayesian source model of data was used by \cite{dong2009truth}, but was limited
in scope to figuring out the evolution over time of the data value.

%

Recent work has also focused on the metrics to use to evaluate data cleaning techniques
\cite{dasu2012statistical}. In this work, we focus on evaluating our method against
the ground truth (when the ground truth is known), and user studies (when the ground truth
is not known). 

While \bw\ uses crowdsourcing to evaluate the accuracy of the proposed
clean tuple alternatives for the experiments on real world datasets,
there are other systems that try to use the crowd for cleaning the
data itself.  X.Chu \etal\ \cite{ppKATARA} clean the database tuples
by discovering patterns that overlap with Knowledge Base(KB)s like
Yago and validating the top-k candidates using the crowd.  J.Wang
\etal\ \cite{crowder} perform entity resolution (which is to identify
several values corresponding to the same entity value) using
crowdsourcing.  They reduce the complexity of the number of HIT(Human
Intelligence Task)s generated by clustering them into several bins so
that a set of pairs can be resolved at a time as against evaluating
one pair at a time.
Y.Zheng \etal\ \cite{qasca} pick a set of $k$ questions to be included in the HITs for the human workers out of a total set of $n$ questions using estimates on 
the expected increase in the answer quality 
by assigning those questions to the crowd. Crowdsourcing to perform data cleaning may be infeasible in the context of Big Data cleaning targeted by \bw\ . However, suggestions from the crowd 
can be used to provide cleaner master data from which \bw\ learns the Bayes network.

The query rewriting part of this work is inspired by the QPIAD system \cite{qpiad2009},
but significantly improves upon it.
QPIAD performed query rewriting over incomplete databases using approximate functional
dependencies (AFD), and only cleaned data with null values, not \emph{wrong} values.

\section{BayesWipe Overview}

\bw\ views the data cleaning problem as a statistical
inference problem over the structured text data. Let $\mathcal{D} =
\{T_1,...,T_n\}$ be the input structured data which contains a
number of corruptions. $T_i \in \mathcal{D}$ is a tuple with $m$ attributes
$\{A_1,...,A_m\}$ which may have one or more corruptions in its
attribute values. Given a candidate replacement set $\mathcal{C}$
for a possibly corrupted tuple $T$ in $\mathcal{D}$, we can clean the database
by replacing $T$ with the candidate clean tuple $T^* \in \mathcal{C}$ that has the
maximum $\textbf{Pr}(T^*|T)$. Using Bayes rule (and dropping the common denominator),
we can rewrite this to
\begin{align}
T^*_{best} &= \arg\max[\textbf{Pr}(T|T^*)\textbf{Pr}(T^*)] \label{eq:offline}
\end{align}
By replacing $T$ with $T^*_{best}$, we get a deterministic database. If we wish to create a probabilistic database (PDB), we don't take an $\arg\max$ 
over the $\textbf{Pr}(T^*|T)$, instead we store the entire distribution over the $T^*$ 
in the resulting PDB.

For online query processing we take the user query $Q^*$, and find the relevance score
of a tuple $T$ as
\begin{align}
\mathit{Score}(T) &= \sum_{T^* \in \mathcal{C}} 
  \underbrace{\textbf{Pr}(T^*)}_\textrm{source model}
  \underbrace{\textbf{Pr}(T|T^*)}_\textrm{error model}
  \underbrace{R(T^*|Q^*)}_\textrm{relevance} \label{eq:basic2}
\end{align}

\noindent
In this work, we used a binary relevance model, where $R$ is 1 if $T^*$ is relevant to the user's
query, and 0 otherwise. Note that $R$ is the relevance of the query $Q^*$ to the candidate clean tuple $T^*$
and not the observed tuple $T$. This allows the query rewriting phase of \bw\, which aims to
retrieve tuples with the highest $Score(.)$ to achieve the non-lossy effect of using a PDB without
explicitly rectifying the entire database.

\smallskip
\noindent\textbf{Architecture:}
\label{sec:archicture}
\label{sec:def}

\begin{figure}[t]
\begin{center}
\includegraphics[keepaspectratio, width=260pt]{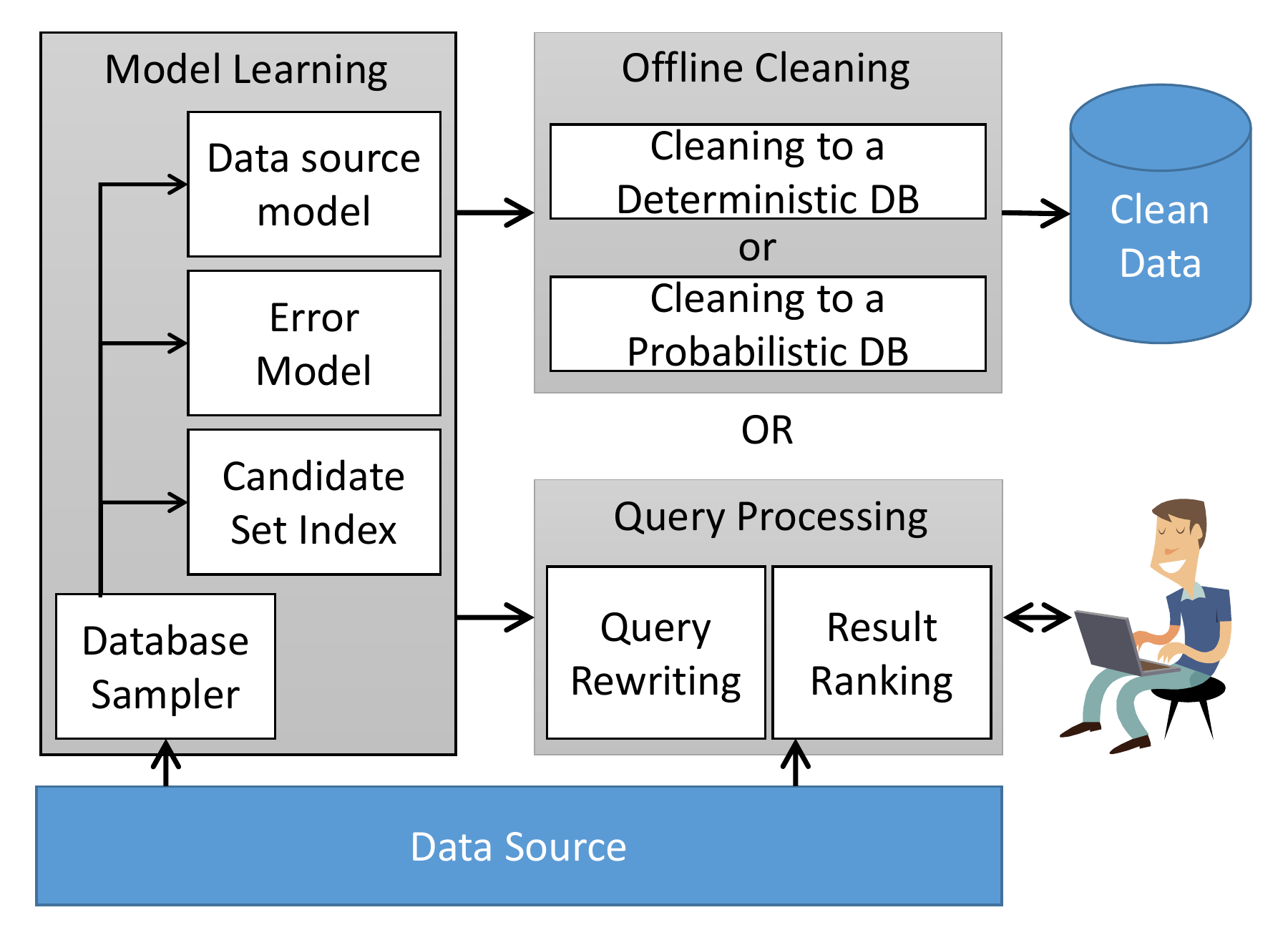}
\caption[The Architecture of BayesWipe]{The architecture of \bw. Our framework learns both data source model and error
model from the raw data during the model learning phase. It can perform offline cleaning
or query processing to provide clean data.}
\mvp\mvp
\label{fig-sys-architecture}
\end{center}
\end{figure}
Figure~\ref{fig-sys-architecture} shows the system architecture for \bw.
During the model learning phase (Section~\ref{sec:model}), we first obtain a sample database
by sending some queries to the database. On this sample data, we learn the generative model
of the data as a Bayes network (Section~\ref{sec:generativemodel}).
In parallel, we define and learn an error model
which incorporates common kinds of errors (Section \ref{sec:errormodel}).
We also create an index to quickly propose candidate $T^*$s.

We can then choose to do either offline cleaning (Section~\ref{sec:offline}) or
online query processing (Section~\ref{sec:query-rewriting}), as per the
scenario.
In the offline cleaning mode, we iterate over all the tuples in the database
and clean them one by one. We can choose whether to store the resulting 
cleaned tuple in a deterministic database (where we store only the $T^*$ with the maximum 
posterior probability) or probabilistic database (where we store the entire distribution 
over the $T^*$). 
In the online query processing mode, we obtain a query from the user,
and do query rewriting in order to find a set of queries that are likely to retrieve
a set of highly relevant tuples. We execute these queries and re-rank the results, and then
display them.

\begin{algorithm}[ht]
  \caption{The algorithm for offline data cleaning}
  \label{algo:offline-algorithm}
  \KwIn{$D$, the dirty dataset.}

  $BN$ $\leftarrow$ Learn Bayes Network ($D$)\\
  \ForEach{Tuple $T \in D$}{
     $\mathcal{C} \leftarrow $ Find Candidate Replacements ($T$)\\
     \ForEach{Candidate $T^* \in \mathcal{C}$}
     {
        $P(T^*) \leftarrow $ Find Joint Probability ($T^*, BN$)\\
        $P(T|T^*) \leftarrow $ Error Model ($T, T^*$)\\
     }
     $T \leftarrow \arg\displaystyle\max_{T^* \in \mathcal{C}} P(T^*)P(T|T^*)$
  }
\end{algorithm}
\begin{algorithm}[h]
  \caption{Algorithm for online query processing.
    }
  \label{algo:online-algorithm}
  \KwIn{$D$, the dirty dataset}
  \KwIn{$Q$, the user's query}

  $S \leftarrow$ Sample the source dataset $D$\\
  $BN$ $\leftarrow$ Learn Bayes Network ($S$)\\
  $ES$ $\leftarrow$ Learn Error Statistics ($S$)\\
  $R \leftarrow$ Query and score results ($Q, D, BN$)\\
  $ESQ \leftarrow$ Get expanded queries ($Q$)\\

  \ForEach{Expanded query $E \in ESQ$}
  {
      $R \leftarrow R\, \cup $ Query and score results ($E, D, BN$)\\
      $RQ \leftarrow RQ\, \cup $ Get all relaxed queries ($E$)\\
  }
  $Sort(RQ)$ by expected relevance, using $ES$\\
  \While{top-$k$ confidence not attained}
  {
      $B \leftarrow$ Pick and remove top $RQ$\\
      $R \leftarrow R\, \cup $ Query and score results ($B, D, BN$)\\
  }
  $Sort(R)$ by score\\
  \Return{R}

\end{algorithm}

In Algorithms~\ref{algo:offline-algorithm} and \ref{algo:online-algorithm}, we present
the overall algorithm for \bw. In the offline mode, we show how we iterate over all the tuples
in the dirty database, $D$ and replace them with cleaned tuples. In the query processing mode,
the first three operations are performed offline, and the remaining operations
show how the tuples are efficiently retrieved from the database, ranked and displayed to the user.

\section{Model Learning}
\label{sec:approach}
\label{sec:model}

\begin{figure}[t]
\centering
\begin{subfigure}[b]{0.23\textwidth}
    \centering
    \includegraphics[trim = 10mm 10mm 10mm 0mm, clip, width=\textwidth, keepaspectratio]{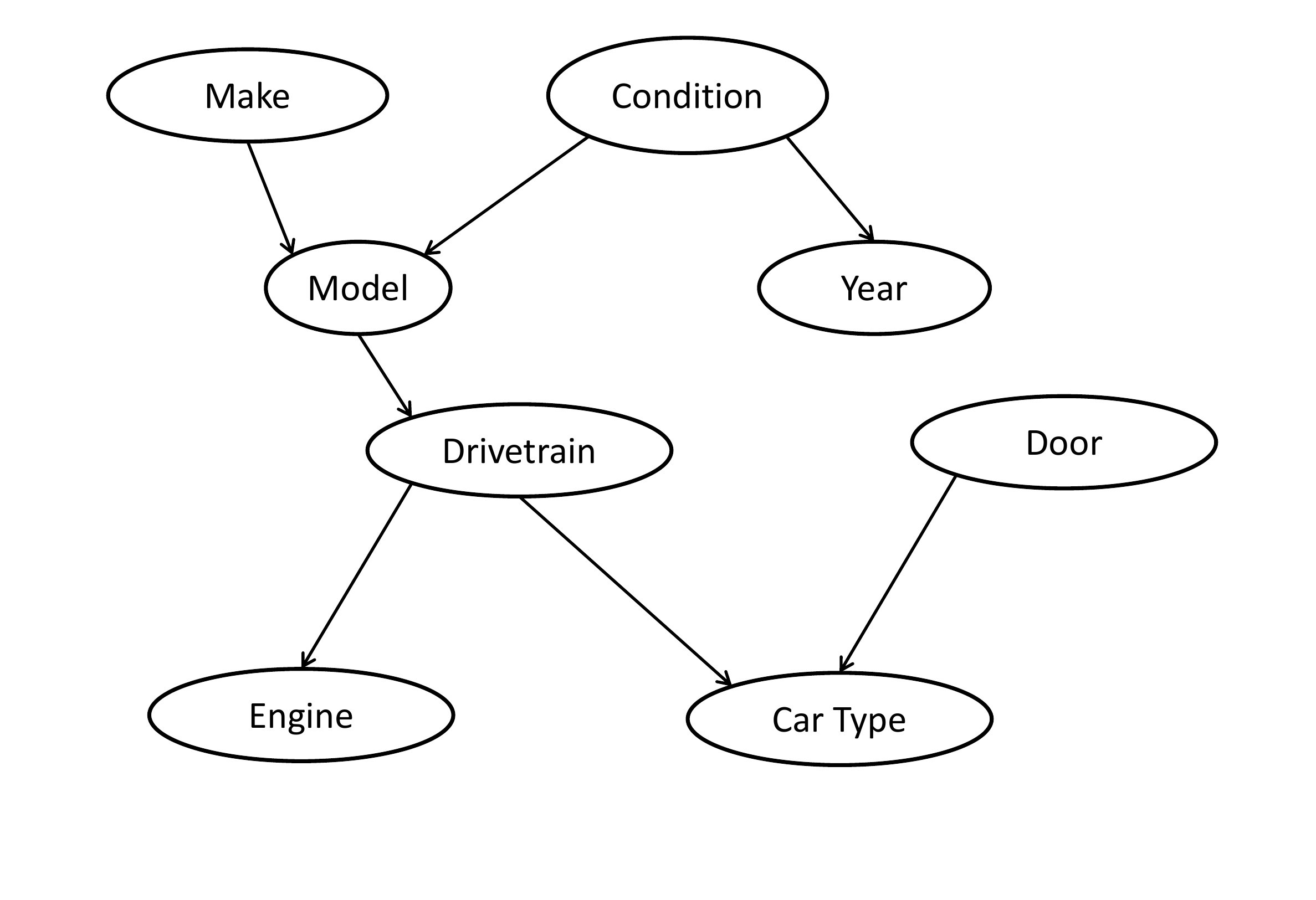}
    \caption{Auto dataset}
    \label{fig:bayes-network}
\end{subfigure}
~
\begin{subfigure}[b]{0.23\textwidth}
    \centering
    \includegraphics[trim = 10mm 10mm 10mm 7mm, clip, width=\textwidth, keepaspectratio]{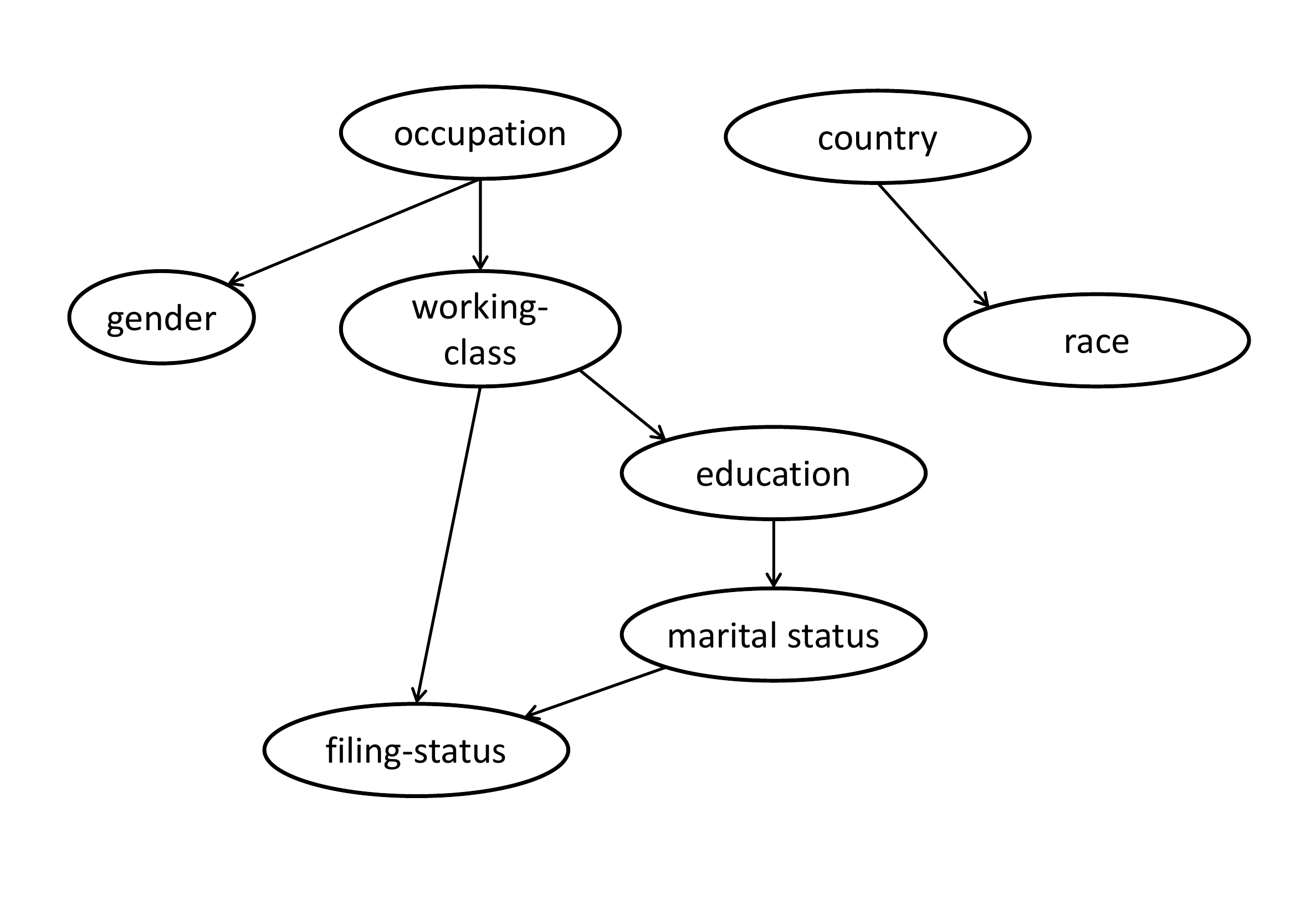}
    \caption{Census dataset}
    \label{fig:bayes-census}
\end{subfigure}
\caption{The learned Bayes networks}
\label{fig:bayes-structures}
\end{figure}

This section details the process by which we estimate the components of Equation~\ref{eq:basic2}:
the data source model ${\textbf{Pr}(T^*)}$ and the error model ${\textbf{Pr}(T|T^*)}$

\subsection{Data Source Model}
\label{sec:generativemodel}

The data that we work with can have dependencies among various attributes (e.g., a car's
{\emph{engine}} depends on its {\emph{make}}). Therefore, we represent the
data source model as a Bayes network, since it naturally captures
relationships between the attributes via structure learning and infers
probability distributions over values of the input tuples.

Constructing a Bayes network over ${\mathcal{D}}$ requires two steps: first, the induction
of the graph structure of the network, which encodes the conditional independences between
the $m$ attributes of ${\mathcal{D}}$'s schema; and second, the estimation of the parameters
of the resulting network. The resulting model allows us to compute probability distributions
over an arbitrary input tuple $T$. 

Whenever the underlying patterns in the source database changes, we have to learn the
structure and parameters of the Bayes network again. In our scenario, we observed that
the structure of a Bayes network of a given dataset remains constant
with small perturbations, but the parameters (CPTs) change more frequently.
As a result, we spend a larger amount of time learning the structure
of the network with a slower, but more accurate tool, Banjo \cite{banjo}.
Figures \ref{fig:bayes-network}
and \ref{fig:bayes-census} show automatically learned structures for two data domains.
The learned structure seems to be intuitively correct, since the nodes that are connected
(for example, `country' and `race' in Figure~\ref{fig:bayes-census}) are expected to
be highly correlated\footnote{Note that the direction of the arrow in a Bayes network
does not necessarily determine causality, see Chapter 14 from Russel and Norvig \cite{russell2010artificial}.}.

Then, given a learned graphical structure ${\mathcal{G}}$ of ${\mathcal{D}}$, we can estimate the conditional probability tables (CPTs) that parameterize each node in ${\mathcal{G}}$ using a
faster package called Infer.NET \cite{InferNET10}. This process of inferring the parameters
is run offline, but more frequently than the structure learning.

Once the Bayesian network is constructed, we can infer the joint
distributions for arbitrary tuple $T$, which can be decomposed to the
multiplication of several marginal distributions of the sets of random
variables, conditioned on their parent nodes
depending on ${\mathcal{G}}$.

\subsection{Error Model}
\label{sec:errormodel}
Having described the data source model, we now turn to the estimation
of the error model ${\textbf{Pr}}(T|T^*)$ from noisy data.
There are many types of errors that can occur in data. We focus on the most
common types of errors that occur in data that is manually entered by na\"ive
users: typos, deletions, and substitution of one word with another. We also
make an additional assumption that error in one attribute does not affect the
errors in other attributes. This is a reasonable assumption to make,
since we are allowing the data itself to have dependencies between attributes,
while only constraining the error process to be independent across attributes.
With these assumptions, we are able to come up with a simple and
efficient error model, where we combine the three types of errors using a
maximum entropy model.

Given a set of clean candidate tuples
$\mathcal{C}$ where $T^* \in \mathcal{C}$, our error model
${\textbf{Pr}}(T|T^*)$ essentially measures how clean $T$ is, or in
other words, how similar $T$ is to $T^*$.

\smallskip
\noindent{\bf Edit distance similarity:}
\label{ed}
This similarity measure is used to detect spelling errors.
Edit distance between two strings $T_{A_i}$ and $T^*_{A_i}$ is defined
as the minimum cost of edit operations applied to dirty tuple
$T_{A_i}$ transform it to clean $T^*_{A_i}$. Edit operations include
character-level copy, insert, delete and substitute.
The cost for each operation can be modified as required; in this paper we use the Levenshtein
distance, which uses a uniform cost function. This gives us a distance, which we
then convert to a probability using \cite{ristad1998learning}:
\begin{equation}
f_{ed}(T_{A_i},T_{A_i}^*) = \exp \{-cost_{ed}(T_{A_i},T_{A_i}^*)\}\label{eq:ed2}
\end{equation}
\noindent{\bf Distributional Similarity Feature:}
This similarity measure is used to detect both substitution and omission errors.
Looking at each attribute in isolation
is not enough to fix these errors. We propose a context-based
similarity measure called Distributional similarity ($f_{ds}$), which
is based on the probability of replacing one value with
another under a similar context \cite{li2006exploring}. Formally, for each string $T_{A_i}$
and $T_{A_i}^*$, we have:
\begin{eqnarray}
f_{ds}(T_{A_i},T_{A_i}^*) = \sum_{c\in C(T_{A_i}, T_{A_i}^*)} 
\frac{{\mathbf{Pr}}(c|T_{A_i}^*){\mathbf{Pr}}(c|T_{A_i}){\mathbf{Pr}}(T_{A_i})}{{\mathbf{Pr}}(c)}\label{eq:6}
\end{eqnarray}
where $C(T_{A_i}, T_{A_i}^*)$ is the context of a tuple attribute value, which is a set of attribute
values that co-occur with both $T_{A_i}$ and $T_{A_i}^*$.
${\mathbf{Pr}}(c|T_{A_i}^*) = (\#(c,T_{A_i}^*)+\mu)/\#(T_{A_i}^*)$ is the probability that a context
value $c$ appears given the clean attribute $T_{A_i}^*$ in the sample database. Similarly,
$P(T_{A_i}) = \#(T_{A_i})/\#tuples$ is the probability that a dirty attribute value appears in the
sample database. We calculate ${\mathbf{Pr}}(c|T_{A_i})$ and ${\mathbf{Pr}}(T_{A_i})$ in the same way.
To avoid zero estimates for attribute values that do not appear in the database sample, we use
Laplace smoothing factor $\mu$.

\smallskip
\noindent{\bf Unified error model:}
\label{unify}
\todo{Reframe:} In practice, we do not know beforehand which kind of error has
occurred for a particular attribute; we
need a unified error model which can accommodate all three types of
errors (and be flexible enough to accommodate more errors when
necessary). For this purpose, we use the well-known maximum entropy
framework \cite{berger1996maximum} to leverage both the
similarity measures, (Edit distance $f_{ed}$ and
distributional similarity $f_{ds}$). For each attribute of the input tuple $T$ and
$T^*$, we have the unified error model ${\mathbf{Pr}}(T|T^*)$ given by:
\begin{align}
\frac{1}{Z} \exp \left\{ \alpha \sum_{i=1}^{m} f_{ed}(T_{A_i}, T_{A_i}^*) 
 +\beta \sum_{i=1}^{m} f_{ds}(T_{A_i}, T_{A_i}^*) \right\} \label{eq:comboerr}
\end{align}
where $\alpha$ and $\beta$ are the weight of each feature, $m$ is the number of attributes in the tuple.
The normalization factor is $Z = \sum_{T^*} \exp \left\{ \sum_{i}  \lambda_i f_{i}(T^*, T)\right\}$.

\subsection{Finding the Candidate Set}
\label{sec:candidate}

The set of candidate tuples, $\mathcal{C}(T)$ for a given tuple $T$ are the
possible replacement tuples that the system considers as possible corrections to $T$. The
larger the set $\mathcal{C}$ is, the longer it will take for the system to
perform the cleaning. If $\mathcal{C}$ contains many unclean tuples, then the
system will waste time scoring tuples that are not clean to begin with.

An efficient approach to finding a reasonably clean $\mathcal{C}(T)$ is to
consider the set of all the tuples in the sample database that differ from $T$
in not more than $j$ attributes. In order to find $\mathcal{C}(T)$ that satisfies this, conceptually, we 
have to iterate over every tuple $t$ in the sample database $D$, comparing it to the tuple $T$ and checking how many 
attributes it differs in. This operation can take $\mathcal{O}(n)$ time, where $n$ is the number of tuples in the sample database.
Even with $j=3$, the na\"ive approach of constructing $\mathcal{C}$ from the sample database directly
is too time consuming, since it requires one to go through the sample database in its entirety once
for every result tuple encountered. To make this process faster, we create indices
over $(j+1)$ attributes because searching through indices reduces the number of comparisons required to
compute $\mathcal{C}(T)$.
If any candidate tuple $T^*$ differs from $T$ in less than or equal to $j$ attributes, then it will
be present in at least one of the indices, since we created $j+1$ of them (pigeon hole principle).
These $j+1$ indices are created over those attributes that have the highest cardinalities,
such as {\sf Make} and {\sf Model} (as opposed to attributes like {\sf Condition} and
{\sf Doors} which can take only a few values).
This ensures that the set of tuples returned from the index  would be small in number.

For every possibly dirty tuple $T$ in the database,
we go over each such index and find all the tuples that match the corresponding attribute.
The union of all these tuples is then
examined and the candidate set $\mathcal{C}$ is constructed by keeping only those tuples from this
union set that do not differ from $T$ in more than $j$ attributes.
Thus we can be sure that by using this method, we have obtained the entire set
$\mathcal{C}$ \footnote{There is a small possibility that the true tuple $T^*$
is not in the sample database at all. This probability can be reduced by
choosing a larger sample set. In future work, we will expand the strategy of
generating $\mathcal{C}$ to include all possible $k$-repairs of a tuple.}.

\vspace{-0.1in}
\section{Offline cleaning}
\label{sec:offline}

\subsection{Cleaning to a Deterministic Database}
\label{sec:detdb}
  In order to clean the data \emph{in situ}, we first use the techniques of the previous section to
  learn the data source model, the error model and create the index. Then, we iterate over all
  the tuples in the database and use Equation~\ref{eq:offline} to find the $T^*$ with the best score.
  We then replace
  the tuple with that $T^*$, thus creating a deterministic database using the offline mode of \bw. 

Computing ${\mathbf{Pr}}(T^*){\mathbf{Pr}}(T|T^*)$ is very fast. Even though we do a Bayesian inference for ${\mathbf{Pr}}(T^*)$, 
the tuple has all the values specified, so the inference ends up being a simple multiplication over the CPTs of the Bayes network, and is very cheap. ${\mathbf{Pr}}(T|T^*)$ 
involves simple edit distance and distributional similarity calculations all of which involve simple arithmetic operations and lookups devoid of Bayesian inference.

  Recall from Section~\ref{sec:errormodel} that there are parameters in the
  error model called $\alpha$ and $\beta$, which need to be set. Interestingly,
  in addition to controlling the relative weight given to the various features in the error model,
  these parameters can be used to control overcorrection by the system.

\noindent  {\bf{Overcorrection}:} Any data cleaning system is vulnerable to
  overcorrection, where a legitimate tuple is modified by the system to an
  unclean value. Overcorrection can have many causes. In a traditional,
  deterministic system, overcorrection can be caused by erroneous rules
  learned from infrequent data. For example, certain makes of cars are all
  owned by the same conglomerate ({\sf{GM}} owns {\sf{Chevrolet}}). In a
  misguided attempt to simplify their inventory, a car salesman might list all
  the cars under the name of the conglomerate. This may provide enough
  support to learn the wrong rule ({\sf{Malibu}} $\to$ {\sf{GM}}).

Typically, once an erroneous rule has been learned, there is no way to correct it or ignore it without a lot of oversight from domain experts. However, \bw\ provides a way to regulate the amount of overcorrection in the system with the help of a `degree of change' parameter. Without loss of generality, we can rewrite Equation~\ref{eq:comboerr} to the following:
  \begin{align*}
    {\mathbf{Pr}}(T|T^*)
      &=\frac{1}{Z} \exp \bigg\{
        \gamma \Big(
          \delta \sum_{i=1}^{m} f_{ed}(T_{A_i}, T_{A_i}^*)
           \\
          &\qquad\qquad+(1 - \delta) \sum_{i=1}^{m} f_{ds}(T_{A_i}, T_{A_i}^*)
        \Big)
      \bigg\}
  \end{align*}
Since we are only interested in their relative weights, the parameters $\alpha$ and $\beta$ have been replaced by $\delta$ and $(1-\delta)$ with the help of a normalization constant, $\gamma$. This parameter, $\gamma$, can be used to modify the degree of variation in ${\mathbf{Pr}}(T|T^*)$.  High values of $\gamma$ imply that small differences in $T$ and $T^*$ cause a larger difference in the value of ${\mathbf{Pr}}(T|T^*)$, causing the system to give higher scores to the original tuple (compared to a modified tuple).

{\bf{Example}:} Consider the following fragment from the database. The first tuple is a very frequent tuple in the database, the second one is an erroneous tuple, and the third tuple is an infrequent, correct tuple. The `true' correction of the second tuple is the third tuple. The ${\mathbf{Pr}}(T^*)$ values shown reflect the values that the data source model might predict for them, roughly based on the frequency with which they occur in the source data.

 \smallskip
 \noindent
\begin{tabular}{|>{}m{0.1in} |>{}m{0.35in} | >{}m{0.32in}| >{}m{0.3in} |>{}m{0.4in} | >{}m{0.53in} |>{}m{0.33in} |}
   \hline
   Id & Make  & Model  & Type  & Engine & Condition & $P(T^*)$ \\\hline\hline
   1  & Honda & Civic  & Sedan & V4    & New   & 0.400  \\\hline
   2  & Honda & Z4     & Sedan & V6    & New   & 0.001  \\\hline
   3  & BMW   & Z4     & Sedan & V6    & New   & 0.005  \\\hline
 \end{tabular}

 \smallskip
 A proper data cleaning system will correct tuple 2 to tuple 3, and not modify any of the others. However, if incorrect rules (for example, {\sf{Z4}} $\to$ {\sf{Honda}}) were learned, there could be overcorrection, where tuple 3 is modified to tuple 2.

 On the other hand, \bw\ handles this situation based on the value of $\gamma$. Looking at tuple 3 (which is a clean tuple), suppose the candidate replacement tuples for it are also tuples 1, 2 and 3. In that case, the situation may look like the following:

 \smallskip
 \noindent
 \begin{tabular}{|c|r||r|r||r|r|}
   \hline
      \multicolumn{2}{|c||}{ } 
    & \multicolumn{2}{|c||}{low $\gamma$} 
    & \multicolumn{2}{|c|}{high $\gamma$} 
   \\\hline
     Cd. & 
     $P(T^*)$  & 
     $P(T|T^*)$& 
     score& 
     $P(T|T^*)$&
     score\\\hline\hline

   1  & 0.400       & 0.02   & {\textbf{0.0080}}  & 0.002 & 0.00080  \\\hline
   2  & 0.001     & 0.30   & 0.0003  & 0.030 & 0.00003  \\\hline
   3  & 0.005     & 1.00   & 0.0050  & 1.000 & {\textbf{0.00500}}  \\\hline
 \end{tabular}
 \smallskip

 As we can see, if we choose a low value of $\gamma$, the candidate with
 the highest score is tuple 1, which means an overcorrection will occur.
 However, with higher $\gamma$, the candidate with the highest score is tuple 3
 itself, which means the tuple will not be modified, and overcorrection will
 not occur. On the other hand, if we set $\gamma$ too high, then even
 legitimately dirty tuples like tuple 2 won't get changed, thus the number of
 actual corrections will also be lower.

 To make full use of this capability of regulating overcorrection, we need to be
 able to set the value of $\gamma$ appropriately. In the absence of a training
 dataset (for which the ground truth is known), we can only estimate the best
 $\gamma$ approximately. We do this by finding a value of $\gamma$ for which
 the percentage of tuples modified by the system is equal to the expected
 percentage of noise in the dataset.
\vspace{-0.05in}
\subsection{Cleaning to a Probabilistic Database}
\label{sec-probdb-comparison}
\label{sec:pdb-creation}
We note that many data cleaning approaches --- including the one we described in the previous
sections --- come up with multiple alternatives for the clean version for any given tuple, 
and evaluate their confidence in each of the alternatives. For example,
if a tuple is observed as {`\textsf{Honda, Corolla}'}, two correct alternatives 
for that tuple might be {`\textsf{Honda, Civic}'} and {`\textsf{Toyota, Corolla}'}. In such cases,
where the choice of the clean tuple is not an obvious one, picking the most-likely 
option may lead to the wrong answer. Additionally, if we intend to do further processing 
on the results, such as perform aggregate queries,
join with other tables, or transfer the data to someone else for processing, then storing
the most likely outcome is lossy.

A better approach (also suggested by others \cite{cra-bigdata}) is to store all of the
alternative clean tuples along with their confidence values. Doing this, however, means
that the resulting database will be a probabilistic database (PDB), even when the source
database is deterministic.

 It is not clear upfront whether PDB-based cleaning will have advantages over 
 cleaning to a deterministic database. On the positive side, using a PDB helps reduce
 loss of information arising from discarding all alternatives to tuples that did not 
 have the maximum confidence. On the negative side, PDB-based cleaning increases the 
 query processing cost (as querying PDBs are harder than querying deterministic databases
 \cite{dalvi2004efficient}).

 Another challenge is one of presentation: users usually assume that they are dealing
 with a deterministic source of data, and presenting all alternatives to them can be
 overwhelming to them.
 In this section, and in the associated experiments, we investigate the potential 
 advantages to using the \bw\ system and storing the resulting cleaned data in a
 probabilistic database. For our experiments, we used Mystiq \cite{boulos2005mystiq},
 a prototype probabilistic database system from University of Washington, as the
 substrate.
 In order to create a probabilistic database from the corrections of the input data, we follow the
 offline cleaning procedure described previously in Section~\ref{sec:model}. Instead of 
 storing the most likely $T^*$, we store all the $T^*$s along with their $P(T^*|T)$ values.
 When evaluating the performance of the probabilistic database, we used simple select queries on 
 the resulting database. Since representing the results of a probabilistic database to the user is a 
 complex task, in this paper we focus on showing the XOR representation of the tuple alternatives to the user. The rationale for our
 decision is that
 in a used car scenario, the user will be provided with a URL link to the car through the clickable tuple id and the several alternative clean values for the dirty attributes
are shown within the single tuple returned to the user. 
 As a result, the form of our output is a tuple-disjoint independent database \cite{suciu2005foundations}. This can be better explained with
 an example:
 \begin{table}[ht]
 \caption{Cleaned probabilistic database}
 \begin{center}
 \begin{scriptsize}
 \begin{tabular}{ | c || c | c | c | c | c | c | c | }
   \hline
   TID & Model & Make  & Orig. & Size & Eng. &  Cond. &P\\
   \hline

 \multirow{2}{*}{$t_1$}
     & {\textsf{Civic}}&{\textsf{Honda}}&{\textsf{JPN}} & {\textsf{Mid-size}} & {\textsf{V4}} & {\textsf{NEW}} & 0.6\\
     & {\textsf{Civic}}&{\textsf{Honda}}&{\textsf{JPN}} & {\textsf{Compact}}&{\textsf{V6}}&{\textsf{NEW}} & 0.4\\
 \hline
 \multicolumn{8}{|c|}{...}\\
 \hline
 \multirow{2}{*}{$t_3$}
     & {\textsf{Civic}}&{\textsf{Honda}}&{\textsf{JPN}}&{\textsf{Mid-size}}&{\textsf{V4}}&{\textsf{USED}} &0.9\\
     & {\textsf{Civik}}&{\textsf{Honda}}&{\textsf{JPN}}&{\textsf{Mid-size}}&{\textsf{V4}}&{\textsf{USED}} &0.1\\
   \hline
 \end{tabular}\label{tbl:cleaned-pdb}
 \end{scriptsize}
 \end{center}
 \end{table}

{\bf{Example}:} Suppose we clean our running example of Table~\ref{tbl:exp1}. We will obtain a
 tuple-disjoint independent\footnote{A tuple-disjoint independent 
 probabilistic database is one where every tuple,
 identified by its primary key, is independent of all other tuples. Each tuple is, however, 
 allowed to have multiple alternatives with associated probabilities. In a tuple-independent
 database, each tuple has a single probability, which is the probability of that tuple
 existing.} probabilistic database \cite{suciu2005foundations};
 a fragment of which is shown in Table~\ref{tbl:cleaned-pdb}.
 Each original input tuple ($t_1, t_3$), has been cleaned, and their alternatives are stored
 along with the computed confidence values for the alternatives (0.6 and 0.4 for $t_1$, in this
 example).
  Suppose the user issues a query
 {\textsf{Model = Civic}}. Both options of tuple $t_1$ of the probabilistic database satisfy the 
 constraints of the query. Since there are two valid alternatives to tuple $t_1$ in the result with probabilities $0.6$ and $0.4$,
in order to get a single tuple representation, the matching attributes in the alternatives are shown deterministically whereas the 
unclean attributes like Size, Engine and Condition with several possible clean values are shown as options. 
 Only the first option in tuple $t_3$ matches the query. Thus the XOR result will contain only a single alternative for
 $t_3$ with probability 0.9. It is important to note that in the case of $t_1$, the Mid-size car can be associated with an Eng. value of V4 and a probability of 0.6 respectively. 
The XOR representation doesn't necessarily allow for combining Mid-size with either an Eng. value of V6 or a probability value of 0.4.

The experimental results compare the tuple ids when computing the 
 recall of the method because tuple id provides the URL to the car's web page which can be used to determine a match.
 The output probabilistic relation is shown in Table~\ref{tbl:result-pdb}.

 \begin{table}[ht]
 \caption{Result probabilistic database}
 \begin{center}
 \begin{scriptsize}
\begin{tabular}{|>{}m{0.1in} |>{}m{0.25in} | >{}m{0.24in}| >{}m{0.2in} |>{}m{0.53in} | >{}m{0.25in} |>{}m{0.25in} |>{}m{0.3in} |}
   \hline
   TID & Model & Make  & Orig. & Size & Eng. &  Cond. &P\\
   \hline
    {$t_1$} & {\textsf{Civic}}&{\textsf{Honda}}&{\textsf{JPN}} & {\textsf{Mid-size/Compact}} & {\textsf{V4/V6}} & {\textsf{NEW}} & 0.6/0.4\\
   \hline
    {$t_3$} & {\textsf{Civic}}&{\textsf{Honda}}&{\textsf{JPN}}&{\textsf{Mid-size}}&{\textsf{V4}}&{\textsf{USED}} &0.9\\
   \hline
 \end{tabular}\label{tbl:result-pdb}
 \end{scriptsize}
 \end{center}
 \end{table}

 The interesting fact here is that the result of any query will always be a tuple-independent 
 database. This is because we projected out every attribute except for the tuple-ID, and the 
 tuple-IDs are independent of each other. 

 When showing the results of our experiments, we evaluate the precision and recall of the system.
 Since precision and recall are deterministic concepts, we have to convert the probabilistic 
 database into a deterministic database (that will be shown to the user) prior to computing these
 values. We can do this conversion
 in two ways: (1) by picking only those tuples whose probability is higher than some threshold. 
 We call this method the \emph{threshold based determinization}. (2) by picking the top-$k$ tuples
 and discarding the probability values (\emph{top-$k$ determinization}). The experiment section 
 (Section~\ref{sec:expt-expt}) shows results with both determinizations.

\vspace{-0.1in}
\section[Query rewriting for Online Query Processing]
{Query rewriting for Online Query Processing}
\label{sec:query-rewriting}
In this section we extend the techniques of the previous section so that it can be used in
an online query processing method where the
result tuples are cleaned at query time. Certain tuples
that do not satisfy the query constraints, but are relevant to the
user, need to be retrieved, ranked and shown to the user. The process
also needs to be efficient, since the time that the users are willing to wait
before results are shown to them is very small.
We show our query rewriting mechanisms aimed at addressing both.

We begin by executing the user's query ($Q^*$) on the database. We store the retrieved results, but do not
show them to the user immediately. We then find rewritten queries that are most likely to
retrieve clean tuples. We do that in a two-stage process: we first expand the query to increase the
precision, and then relax the query by deleting some constraints (to increase the recall).
\vspace{-0.05in}
\subsection{Increasing the precision of rewritten queries}
We can improve precision
by adding relevant constraints to the query $Q^*$ given by the user. For example, when a user issues the
query ${\textsf{Model = Civic}}$, we can expand the query to add relevant constraints ${\textsf{Make = Honda,
Country = Japan, Size = Mid-Size}}$. These additions capture the essence of the query --- because they limit
the results to the specific kind of car the user is probably looking for. These expanded structured queries
generated from the user's query are called $ESQ$s.

Each user query $Q^*$ is a select query with one or more attribute-value pairs as constraints. In order to create an $ESQ$, we will have to add highly correlated constraints to $Q^*$.

Searching for correlated constraints to add requires Bayesian inference, which is an expensive operation. Therefore, when searching for constraints to add to $Q^*$, we restrict the search to the union of all the attributes in the Markov blanket \cite{pearl1988probabilistic}. The Markov blanket of an attribute comprises its children, its parents, and its children's other parents. It is the set of attributes whose value being given, the node becomes independent of all other nodes in the network. Thus, it makes sense to consider these nodes when finding correlated attributes.
This correlation is computed using the Bayes Network that was
learned offline on a sample database (recall the architecture of
\bw\ in Figure~\ref{fig-sys-architecture}.)

Given a $Q^*$, we attempt to generate multiple $ESQ$s that maximizes both the relevance of the results
and the coverage of the queries of the solution space.

Note that if there are $m$ attributes, each of which can take $n$ values, then the total number
of possible $ESQ$s is $n^m$. Searching for the $ESQ$ that globally maximizes the
objectives in this space is infeasible; we therefore approximately search for it by performing a heuristic-informed search. Our objective is to create an $ESQ$ with $m$ attribute-value pairs as constraints.We begin with the constraints specified by the user query $Q^*$. We set these as evidence
in the Bayes network, and then query the Markov blanket of these attributes for the attribute-value
pairs with the highest posterior probability given this evidence. We take the top-$k$ attribute-value pairs and append them to $Q^*$ to produce $k$ search nodes, each search node being
a query fragment. If $Q$ has $p$ constraints in it, then the heuristic value of $Q$ is given by
${\textbf{Pr}}(Q)^{m/p}$. This represents the expected joint probability of $Q$ when expanded to
$m$ attributes, assuming that all the constraints will have the same average posterior probability.
We expand them further, until we find $k$ queries of size $m$ with the highest probabilities.
\vspace{-2mm}
\begin{figure}[htb]
\includegraphics[width=0.48\textwidth]{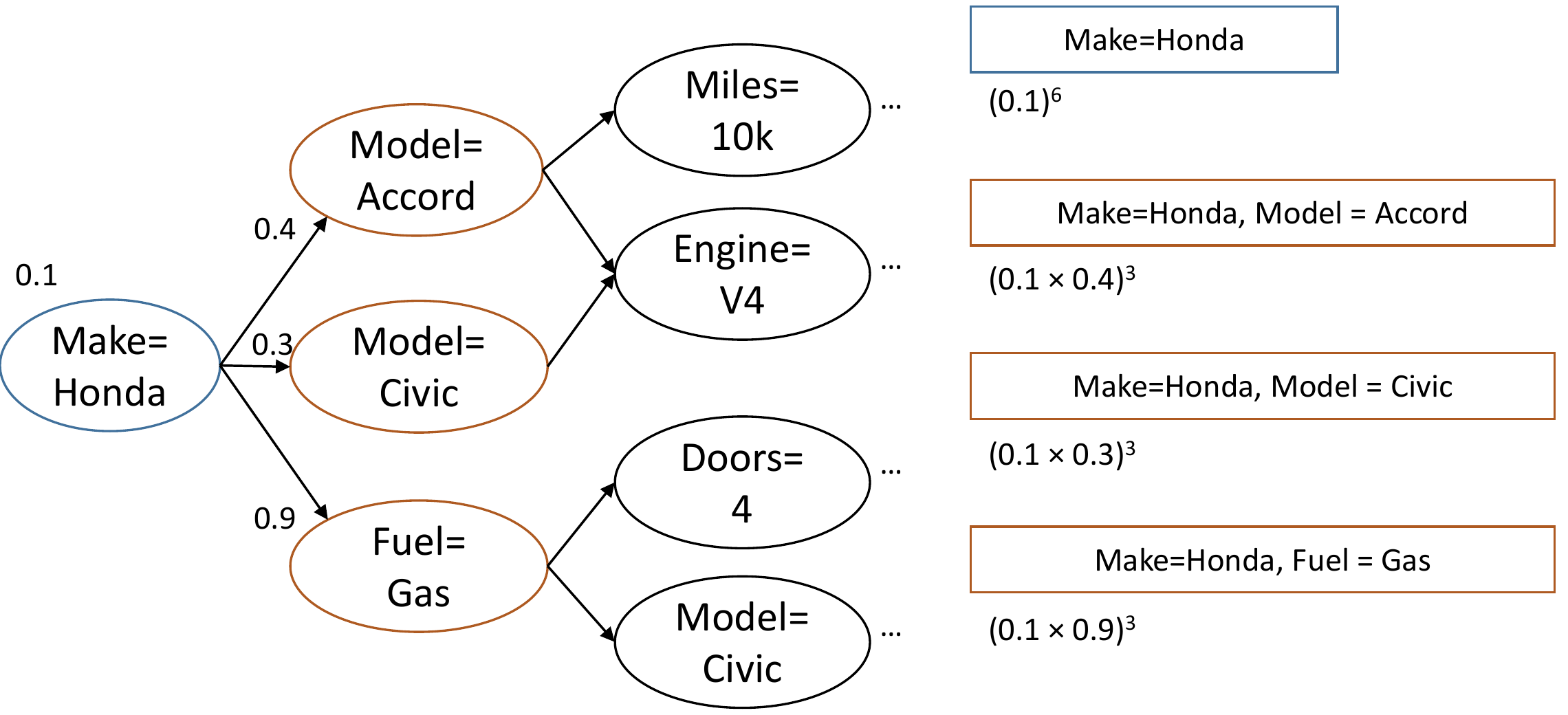}
\mvp
\caption{Query Expansion Example. The tree shows the candidate constraints that can be
added to a query, and the rectangles show the expanded queries with the computed
probability values.}
\label{fig:query-exapansion}
\end{figure}
\vspace{-2mm}

 {\bf{Example}}: In Figure~\ref{fig:query-exapansion}, we show an example of the query expansion. The node
 on the left represents the query given by the user ``Make=Honda''. First, we look at the Markov Blanket
 of the attribute Make, and determine that Model and Condition are the nodes in the Markov blanket.
 we then set ``Make=Honda'' as evidence in the Bayes network and then run an inference over the 
 values of the attribute Model. The two values of the Model attribute with the highest posterior
 probability are Accord and Civic. The most probable values of the Condition attribute are ``new''
 and ``old''. Using each of these values, new queries are constructed and added to the queue.
 Thus, the queue now consists of the 4 queries: ``Make=Honda, Model=Civic'', ``Make=Honda, Model=Accord''
 and ``Make=Honda, Condition=old''. A fragment of these queries are shown in the middle column
 of Figure~\ref{fig:query-exapansion}. We dequeue the highest probability item from the queue
 and repeat the process of setting the evidence, finding the Markov Blanket, and running the inference. We stop when we get the required number of $ESQ$s with a sufficient number of constraints. 
\vspace{-0.05in}
\subsection{Increasing the recall}
Adding constraints to the query causes the precision of the results to increase, but
reduces the recall drastically. Therefore, in this stage, we choose to delete
some constraints from the $ESQ$s, thus generating relaxed queries ($RQ$).
Notice that tuples that have corruptions in the attribute constrained by the user
can only be retrieved by relaxed queries that do not specify a value for
those attributes. Instead, we have to depend on rewritten queries that contain correlated
values in other attributes to retrieve these tuples.
Using relaxed queries can be seen as a trade-off between the recall of the resultset
and the time taken, since
there are an exponential number of relaxed queries for any given $ESQ$. As a result,
an important question is the choice of $RQ$s to execute. We take the approach of generating every possible $RQ$, and then ranking them according to their expected relevance. This operation is performed entirely on the learned error statistics, and is thus very fast.

We score each relaxed query by the \emph{expected relevance} of its result set.
\begin{align*}
Rank(q) = \mathbb{E}\left( \frac{\sum_{T_q} \textit{Score}(T_q|Q^*)}{|T_q|}\right)
\end{align*}
where $T_q$ are the tuples returned by a query $q$,
and $Q^*$ is the user's query. Executing an $RQ$ with a higher rank will have a more beneficial
result on the result set because it will bring in better quality result tuples.
Estimating this quantity is difficult because we do not have
complete information about the tuples that will be returned for any query $q$.
The best we can do, therefore, is to approximate this quantity.

Let the relaxed query be $Q$, and the expanded query that it was relaxed from be $ESQ$. We
wish to estimate ${\mathbb{E}}[P(T|T^*)]$ where $T$ are the tuples returned by $Q$. Using the
attribute-error independence assumption, we can rewrite that as $\prod_{i=0}^m \textbf{Pr}(T.A_i|T^*.A_i)$,
where $T.A_i$ is the value of the $i$-th attribute in T. Since $ESQ$ was obtained by expanding
$Q^*$ using the Bayes network, it has values that can be considered clean for this evaluation.
Now, we divide the $m$ attributes
of the database into 3 classes: (1) The attribute is specified both in $ESQ$ and in $Q$.
In this case, we set ${\textbf{Pr}}(T.A_i|T^*.A_i)$ to 1, since $T.A_i = T^*.A_i$. (2) The attribute
is specified in $ESQ$ but not in $Q$. In this case, we know what $T^*.A_i$ is, but not
$T.A_i$. However, we can generate an average statistic of how often $T^*.A_i$ is erroneous
by looking at our sample database. Therefore, in the offline learning stage, we precompute
tables of error statistics for every $T^*$ that appears in our sample database, and use that
value. (3) The attribute
is not specified in either $ESQ$ or $Q$.
 In this case, we know neither the attribute value in
$T$ nor in $T^*$. We, therefore, use the average error rate of the entire attribute as the
value for ${\textbf{Pr}}(T.A_i|T^*.A_i)$. This statistic is also precomputed during the learning phase.
This product gives the expected rank of the tuples returned by $Q$.
 \begin{figure}[htb]
 \centering
 \includegraphics[width=0.48\textwidth]{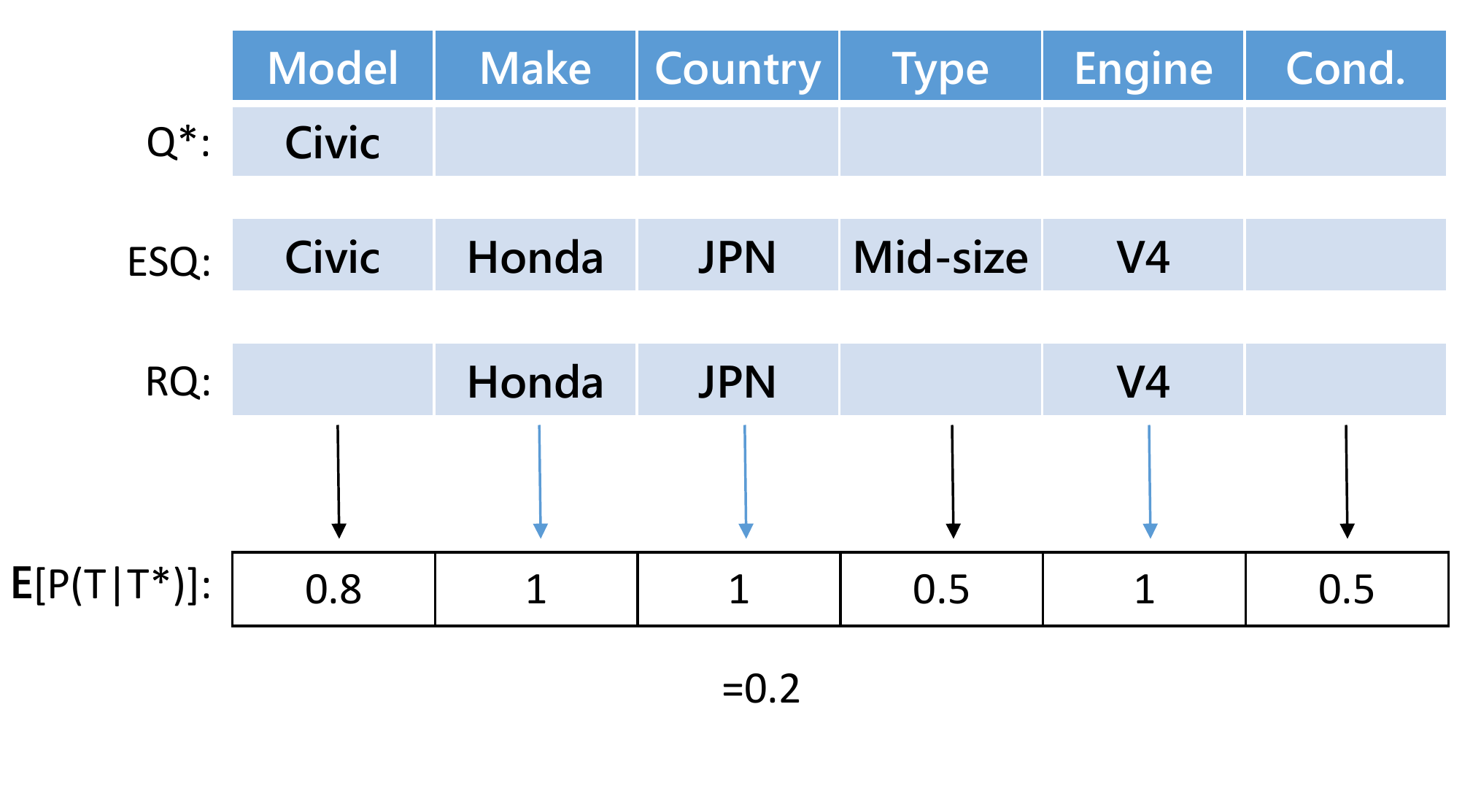}
\mvp
\mvp
 \caption{Query Relaxation Example.}
 \label{fig:query-relaxation}
 \vspace{-0.2in}
 \end{figure}

 \textbf{Example:}
 In Figure~\ref{fig:query-relaxation}, we show an example for finding the probability values of a relaxed query. Assume that the user's query $Q^*$ is ``Civic'', and the $ESQ$ is shown in the second row. For an RQ that removes the attribute values ``Civic'' and ``Mid-Size'' from the $ESQ$, the probabilities are calculated as follows: For the attributes ``Make, Country'' and ``Engine'',
 the values are present in both the $ESQ$ as well as the $RQ$, and therefore, the $P(T|T^*)$ 
 for them is 1. For the attribute ``Model'' and ``Type'', the values are present in $ESQ$ but not in $RQ$, hence the value for them can be computed from the learned error statistics. For example, for ``Civic'', the average value of $P(T|Civic)$ as learned from the sample database (0.8) is used.
 Finally, for the attribute ``Condition'', which is present neither in $ESQ$ nor in $RQ$, we use the average error statistic for that attribute (i.e. the average of $P(T_a|T^*_a)$ for $a$ = ``Condition'' which is 0.5).

 The final value of ${\mathbb{E}}[P(T|T^*)]$  is found from the product of all these attributes as 0.2. This process is very fast because it only involves lookups and multiplication - bayesian inference is not needed.

\vspace{-0.05in}
\subsection{Terminating the process}
We begin by looking at all the $RQ$s in descending order
of their rank. If the current $k$-th tuple in our resultset has a relevance of $\lambda$,
and the estimated rank of the $Q$ we are about to execute is $R(T_q|Q)$, then we
stop evaluating any more queries if the probability ${\mathbf{Pr}}(R(T_q|Q) > \lambda)$ is
less than some user defined threshold ${\mathcal{P}}$. This ensures that we have the true top-$k$
resultset with a probability ${\mathcal{P}}$.

\vspace{-0.1in}
\section{Map-Reduce framework}
\label{sec:mapreduce}
BayesWipe is most useful for big-data related scenarios. BayesWipe has two modes: online and offline. The online mode of BayesWipe already works for big data scenarios by optimising the rewritten queries it issues. Now, we show that the offline mode can also be optimized for a big-data scenario by implementing it as a Map-Reduce application.


So far, BayesWipe-Offline has been implemented as a two-phase, single threaded program. 
In the first phase, the program learns the Bayes network (both structure and parameters), learns the error statistics, and creates the candidate index. Recall from section~\ref{sec:candidate} that we create an index on the attributes of the sample database to speed up the creation of the candidate set of clean tuples; which we refer to as the candidate index. The candidate index is constructed on a set of $j+1$ attributes when the restriction on a candidate clean tuple is to differ from the dirty tuple in not more than $j$ attributes. The attributes in the dirty tuple are compared to the attributes of the tuples in the sample database using the candidate index to generate the set of candidate clean tuples. Note that this candidate index can be constructed on any arbitrary set of $j+1$ attributes present in the sample database. 
In the second phase, the program goes through every tuple in the input database, picks a set of candidate tuples, and then evaluates the $P(T^*|T) P(T^*)$ for every candidate tuple, and replaces $T$ with the $T^*$ that maximises that value. Since the learning is typically done on a sample of the data, it is more important to focus on the second phase for the parallelizing efforts. Later, we will see how the learning of the error statistics can also be parallelized.

\vspace{-0.05in}
\subsection{Simple Approach}
The simplest approach to parallelizing \bw\ is to run the first phase (the learning phase) on a single machine. Then, a copy of the bayes network (structure and CPTs), the error statistics, and the candidate index can be sent to a number of other machines. Each of those machines also receives a fraction of the input data from the dirty database. With the help of the generative model and the input data, it can clean the tuples, and then create the output.

If we express this in Map-Reduce terminology, we will have a pre-processing step where we create the generative and error models. The Map-Reduce architecture will have only mappers, and no reducers. The result of the mapping will be the tuple ${\langle T, T^* \rangle}$.

The problem with this approach is that in a truly big data scenario, the candidate index can become very large. Indeed, as the number of tuples increases, the size of the domain of each attribute also increases (see Figure~\ref{fig:shards-tuples} for 1 shard). Further, the number of different combinations, and the number of erroneous values for each attribute also increase (Figure~\ref{fig:shards-noise}). All of this results in a rather large candidate index. Transmitting and using the entire index on each mapper node is wasteful of both network, memory, (and if swapped out, disk resources). Note that to create a rich and useful data correction system, we have to accommodate a large candidate clean-tuple set, ${\mathcal{C}}(T)$, for every $T$. ${\mathcal{C}}(T)$ roughly tracks the sample database size. If we are unable to shard ${\mathcal{C}}(T)$, then sharding the input data is pointless. In the following section we endeavor to show a strategy where not just the input, but also the index on the candidate set ${\mathcal{C}}(T)$ can be sharded across machines.

\vspace{-0.05in}
\subsection{Improved Approach}
In order to split both the input tuples and the candidate index, we use a two-stage approach. In the first stage, we run a map-reduce that splits the problem into multiple shards, each shard having a small fraction of the candidate index. The second stage is a simple map-reduce that picks the best output from stage 1 for each input tuple.

Stage 1: Given an input tuple $T$ and a set of candidate tuples, the $T^*$s, suppose the candidate index is created on $k$ attributes, $A_1 ... A_k$. We can say that for every tuple $T$, and one of its candidate tuples $T^*$, they will have at least one matching attribute $a_i$ from this set. We can use this common element $a_i$ to predict which shards the candidate $T^*$s might be available in. We therefore, send the tuple $T$ to each shard that matches the hash of the value $a_i$.

In the map-reduce architecture, it is possible to define a `partition' function. Given a mapped key-value pair, this function determines which reducer nodes will process the data. We can use an exact equivalence on each value that the matching attribute can take, $a_i$ as the partition function. However, notice that the number of possible values that $A_1 ... A_k$ can take is rather large. If we na\"ively use $a_i$ as the partition function, we will have to create those many reducer nodes. Therefore, more generally, we hash this value into a fixed number of reducer nodes, using a deterministic hash function. This will then find all candidate tuples that are eligible for this tuple, compute the similarity, and output it.

{\bf{Example}}: Suppose we have tuple $T_1$ that has values $(a_1, a_2, a_3, a_4, a_5)$. Suppose our candidate index is created on attributes $A_1, A_2, A_4$. This means that any candidates $T^*$ that are eligible for this tuple have to match one of the values $a_1, a_2$ or $a_4$. Then the mapper will create the pairs $(a_1, T)$, $(a_2, T)$ and $(a_4,T)$, and send to the reducers. The partition function is the hash of the key - so in this case, the first one will be sent to the reducer number $hash(A1=a1)$, the second will be sent to the reducer numbered $hash(A2=a2)$, and so on. 

In the reducer, the similarity computation and computation of the prior probabilities from the \bw\ algorithm are run. Since each reducer only has a fraction of the candidate index (the part that matches $A1=a1$, for instance), it can hold it in memory and computation is quite fast. Each reducer produces a pair $(T_1, (T_{1n}^*, {\text{score}}))$. Since there are several candidate clean tuples, $n$ is used to identify a specific tuple among those alternatives.

Stage 2: This stage is a simple $\max$ calculation. The mapper does nothing, it simply passes on the key-value pair $(T_1, (T_{1n}^*, {\text{score}}))$ that was generated in the previous Map-Reduce job. Notice that the key of this pair is the original, dirty tuple $T_1$. The Map-Reduce architecture thus automatically groups together all the possible clean versions of $T_1$ along with their scores. The reducer picks the best T* based on the score (using a simple $\max$ function), and outputs it to the database.

\vspace{-4mm}
\begin{figure}[ht]
\centering
\includegraphics[width=0.48\textwidth]{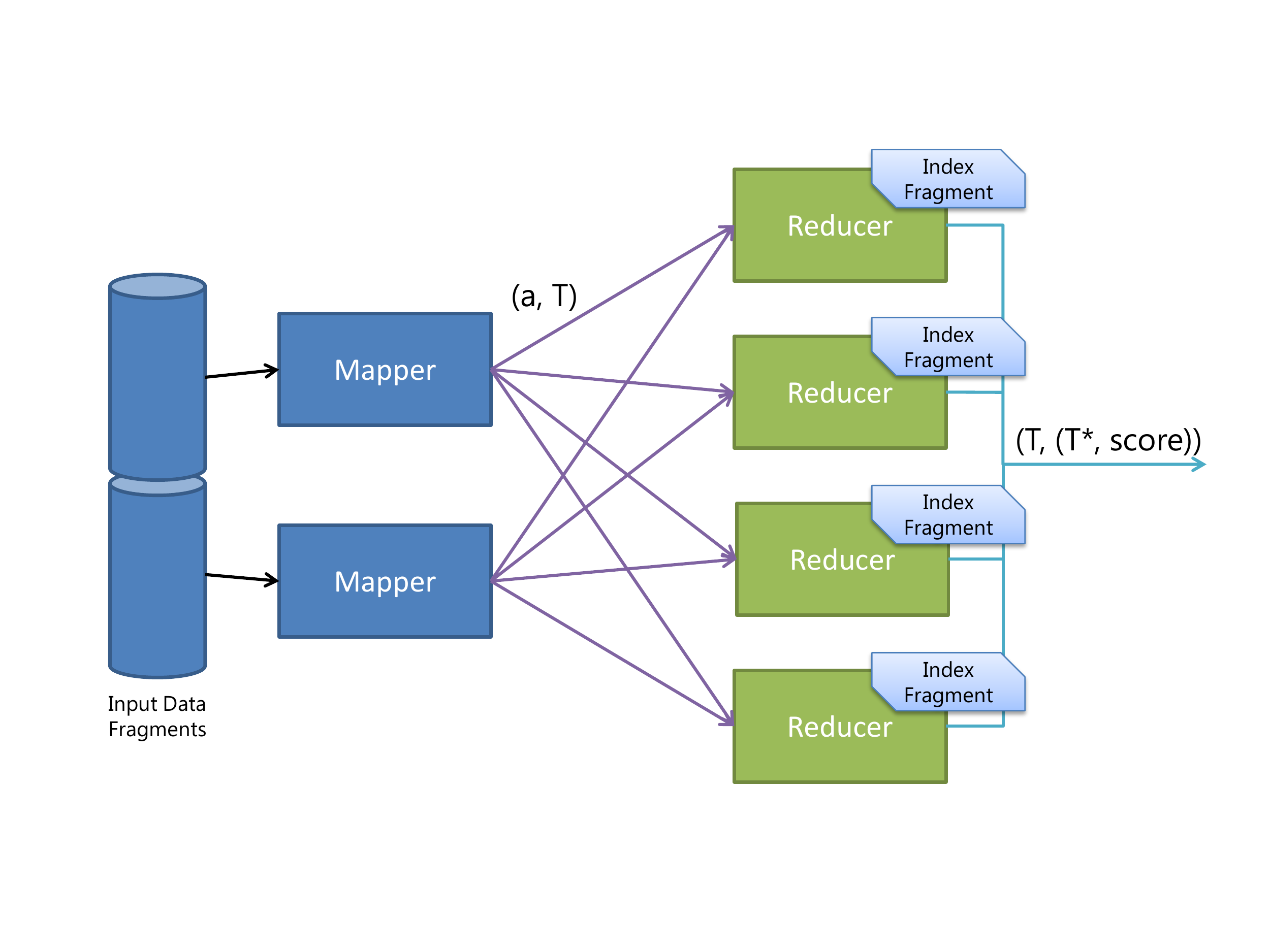}
  \vspace{-0.1in}
\mvp\mvp
  \caption{Stage-1 Map-Reduce Framework for BayesWipe.}
  \mvp \mvp
  \label{fig:mapred}
\end{figure}

\subsection{Results of This Strategy}
In Figure~\ref{fig:shards-tuples} and Figure~\ref{fig:shards-noise} we can see how this map reduce strategy helps in reducing the memory footprint of the reducer. First, we plot the size of the index that needs to be held in each node as the number of tuples in the input increases. The topmost curve shows the size of index in bytes if there was no sharding - as expected, it increases sharply. The other curves show how the size of the index in the one of the nodes varies for the same dataset sizes. From the graph, it can be seen that as the number of tuples increases, the size of the index grows at a lower rate when the number of shards is increased. This shows that increasing the number of reduce nodes is a credible strategy for distributing the burden of the index.

In the second figure (Figure~\ref{fig:shards-noise}), we see how the size of the index varies with the percentage of noise in the dataset. As expected, when the noise increases, the number of possible candidate tuples increase (since there are more variations of each attribute value in the pool). Without sharding, we see that the size of the dataset increases. While the increase in the size of the index is not as sharp as the increase due to the size of the dataset, it is still significant. Once again, we observe that as the number of shards is increased, the size of the index in the shard reduces to a much more manageable value.

Note that a slight downside to this architecture is that a shuffle/reduce of $(T, (T_{n}^*, {\text{score}}))$ is needed in the second stage, and this intermediate data structure can be quite large. While this leads to some network and temporary storage overhead, the primary objective of sharding the expensive computation has been achieved by this architecture.

\vspace{-0.3in}
\label{expt-correctness}
\section{Empirical Evaluation}
\label{sec:experiments}
We quantitatively study the performance of \bw\ in both its modes --- offline,
and online, and compare it against state-of-the-art CFD approaches. 
We present experiments on evaluating  the
approach in terms of the effectiveness of data cleaning, efficiency and
precision of query rewriting.
A demo for the offline cleaning mode of \bw\ can be downloaded from
 {http://bayeswipe.sushovan.de/}.

\vspace{-0.1in}
 \subsection{Datasets}
 \label{sec:expt-data}
 To perform the experiments, we obtained the real data from the web.
 The first dataset is \emph{Used car sales} dataset $D_{car}$ 
 crawled from Google Base. 
Such ``dirty" dataset is referred to as ``$D'_{car}$". 
 The second dataset we used was adapted from the \emph{Census Income}
 dataset $D_{census}$ from the UCI machine learning repository
 \cite{asuncion2007uci}. From the fourteen available attributes, we
 picked the attributes that were categorical in nature, resulting in
 the following 8 attributes: \textsf{working-class, education, marital
   status, occupation, race, gender, filing status. country}. The same
 setup was used for both datasets -- including parameters and
 error features.

 These datasets
 were observed to be mostly clean. We then introduced\footnote{
 We note that the introduction of synthetic errors into 
 clean data  for experimental evaluation purposes is common practice
 in data cleaning research \cite{cong2007improving, bohannon2007conditional}.} 
 three types of noise to the attributes. To add noise to an attribute, we randomly
 changed it either to a new value which is close in terms of string
 edit distance (distance between 1 and 4, simulating spelling errors)
 or to a new value which was from the same attribute (simulating
 replacement errors) or just delete it (simulating deletion errors).
 As we have mentioned before, one of the assumptions of this paper is that
 the error model is a combination of these three kinds of errors, and that the
 errors are independent of each other. By synthetically generating these errors,
 we were able to test our system against a dataset that satisfies the assumption.

 The next dataset tests our system against a real-world scenario where we do not
 control the error process, and thus validates that this assumption was not
 unrealistic.

 To test our system against real-world noise where we do not have any control
 over amount, type or behavior of the noise generation process, we crawled car
 inventory data from the website `cars.com'. We manually verified that the
 data obtained did, in fact, have a reasonable number of inaccuracies, making it
 a suitable candidate for testing our system.

\vspace{-0.1in}
\subsection{Experiments}
\label{sec:expt-expt}
\begin{figure*}[ht]
\centering
\begin{subfigure}{0.31\textwidth}
    \centering
    \includegraphics[width=\textwidth, keepaspectratio]{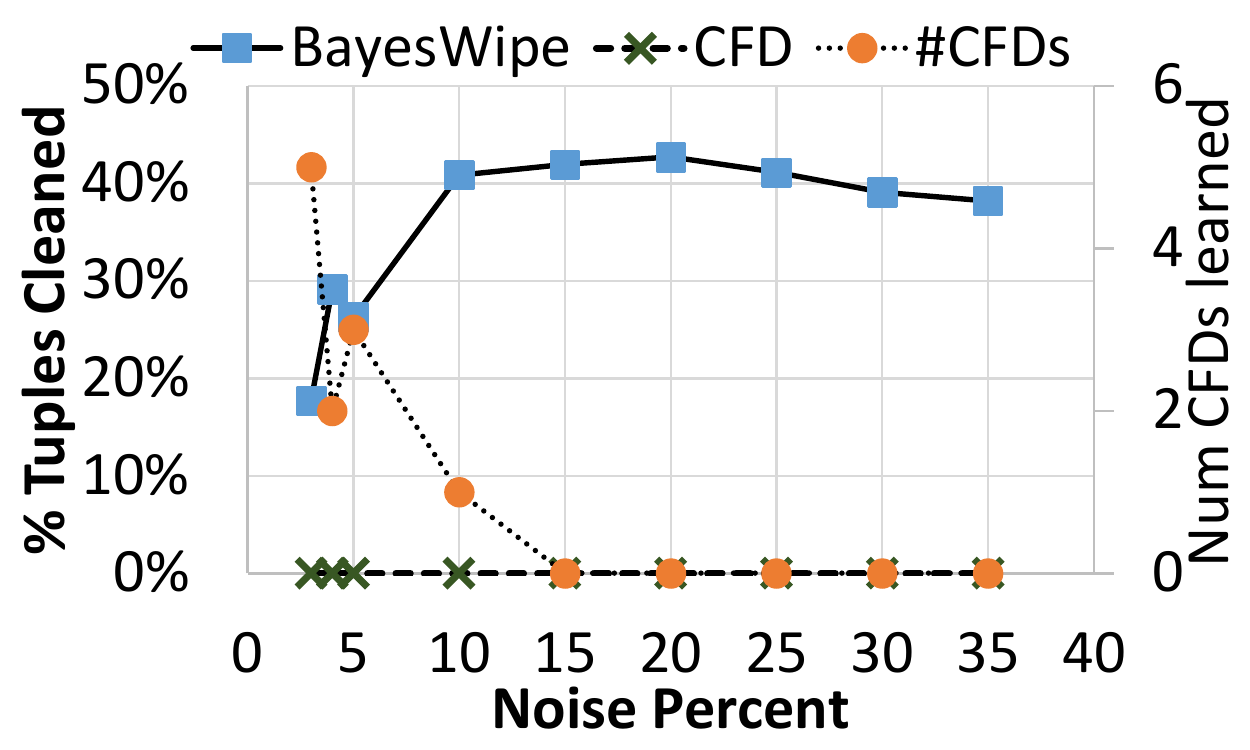}
    \caption{\% performance of \bw\ compared to CFD, for the used-car dataset.}
    \label{fig-offline-clean-vs-cfd}
\end{subfigure}%
\quad
\begin{subfigure}{0.31\textwidth}
    \centering
    \includegraphics[width=\textwidth, keepaspectratio]{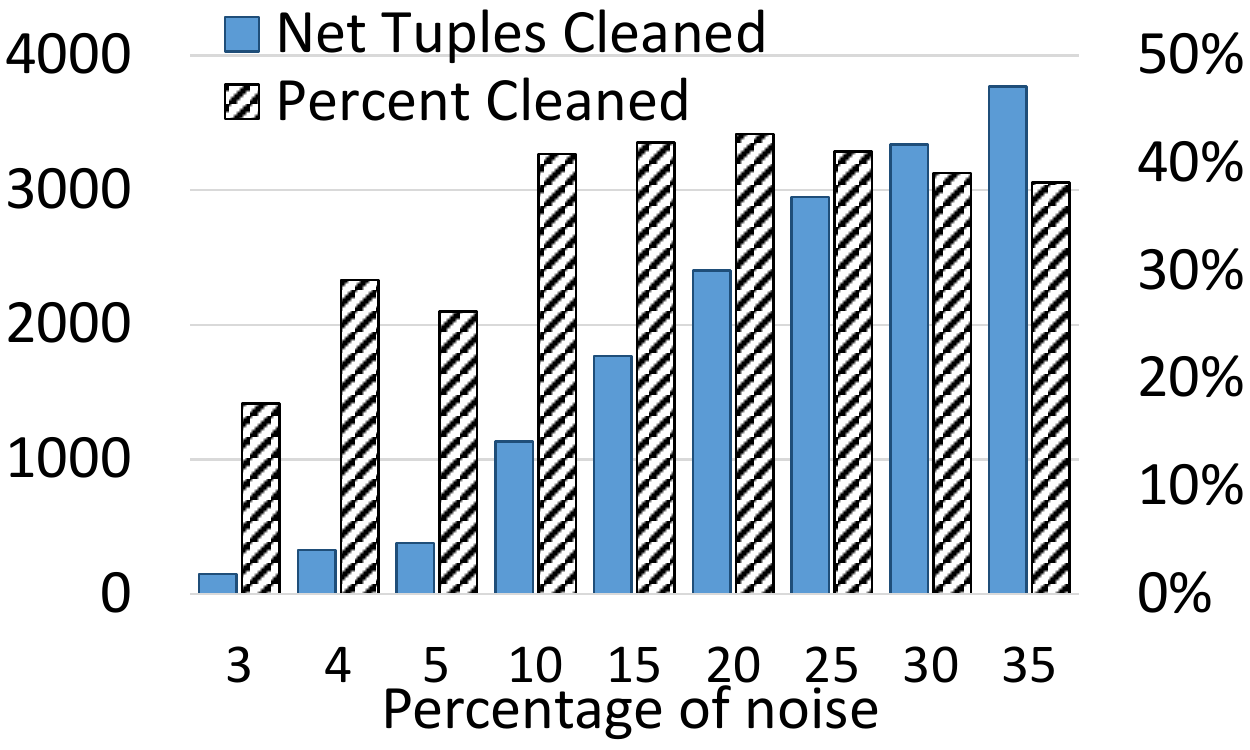}
    \caption{\% net corrupt values cleaned, car database}
    \label{fig-offline-clean-vs-noise}
\end{subfigure}%
\quad
\centering
\begin{subfigure}{0.31\textwidth}
  \includegraphics[keepaspectratio, width=\textwidth,
    clip, trim=0pt 10pt 0pt 0pt]{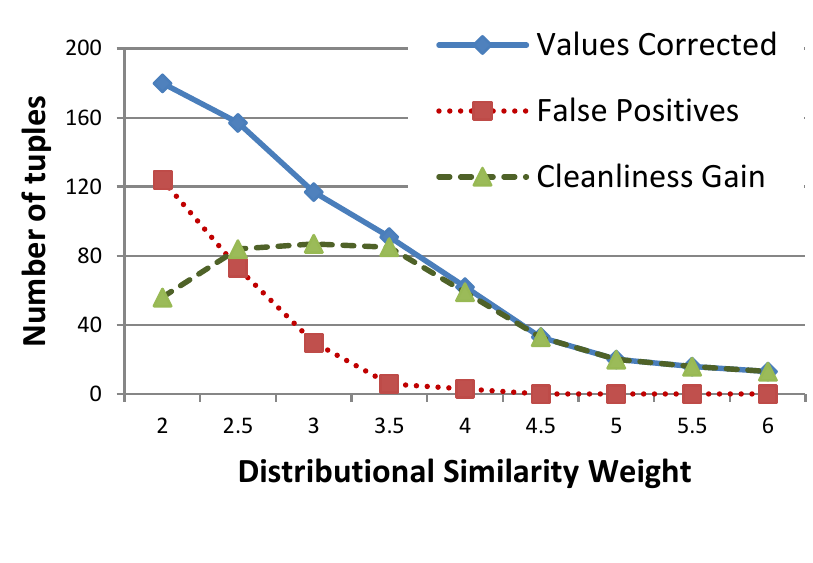}
  \caption{Net corrections vs $\gamma$. (The $x$-axis values show the un-normalized distributional similarity weight, which is simply $\gamma \times 3/5$.)}
\mvp
  \label{fig:gain-vs-parameter}
\end{subfigure}

\caption{Offline cleaning mode of \bw}
\mvp
\label{fig-offline-cleaning}
\end{figure*}

\smallskip\noindent
\textbf{Offline Cleaning Evaluation:}
The first set of evaluations shows the effectiveness of the offline cleaning mode.
In Figure~\ref{fig-offline-clean-vs-cfd}, we compare \bw\ against
CFDs \cite{chiang2008discovering}.

The dotted line that shows the number of CFDs learned from the noisy data
quickly falls to zero, which is not surprising:
CFDs learning was designed with a clean training dataset in mind. Further, the only constraints
learned by this algorithm are the ones that have not been violated in the dataset ---
unless a tuple violates some CFD, it cannot be cleaned. As a result,
the CFD method cleans exactly zero tuples independent of the noise percentage. On the other hand, \bw\ is
able to clean between 20\% to 40\% of the incorrect values. It is interesting to note that the percentage
of tuples cleaned increases initially and then slowly decreases, because for very low
values of noise, there aren't enough errors for the system to learn a reliable error
model from; and at larger values of noise, the data source model learned from the noisy data is of
poorer quality.

 \begin{figure*}[t]
 \centering
 \begin{subfigure}{0.55\textwidth}
   \centering
   \includegraphics[width=\textwidth]{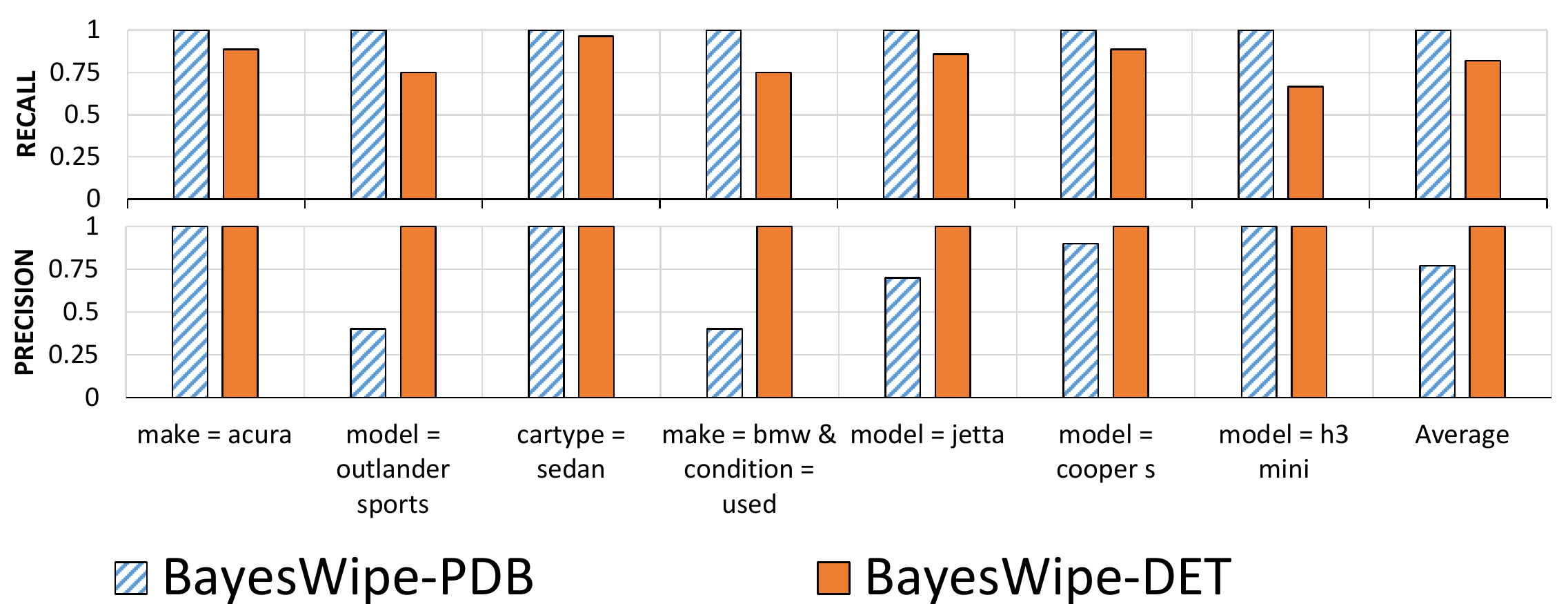}
   \caption{Precision and recall of the PDB method shown against the deterministic method for specific queries.}
   \label{fig:pdb-prec-recall-vs-queries}
 \end{subfigure}
 ~
 \begin{subfigure}{0.20\textwidth}
   \centering
   \includegraphics[width=\textwidth]{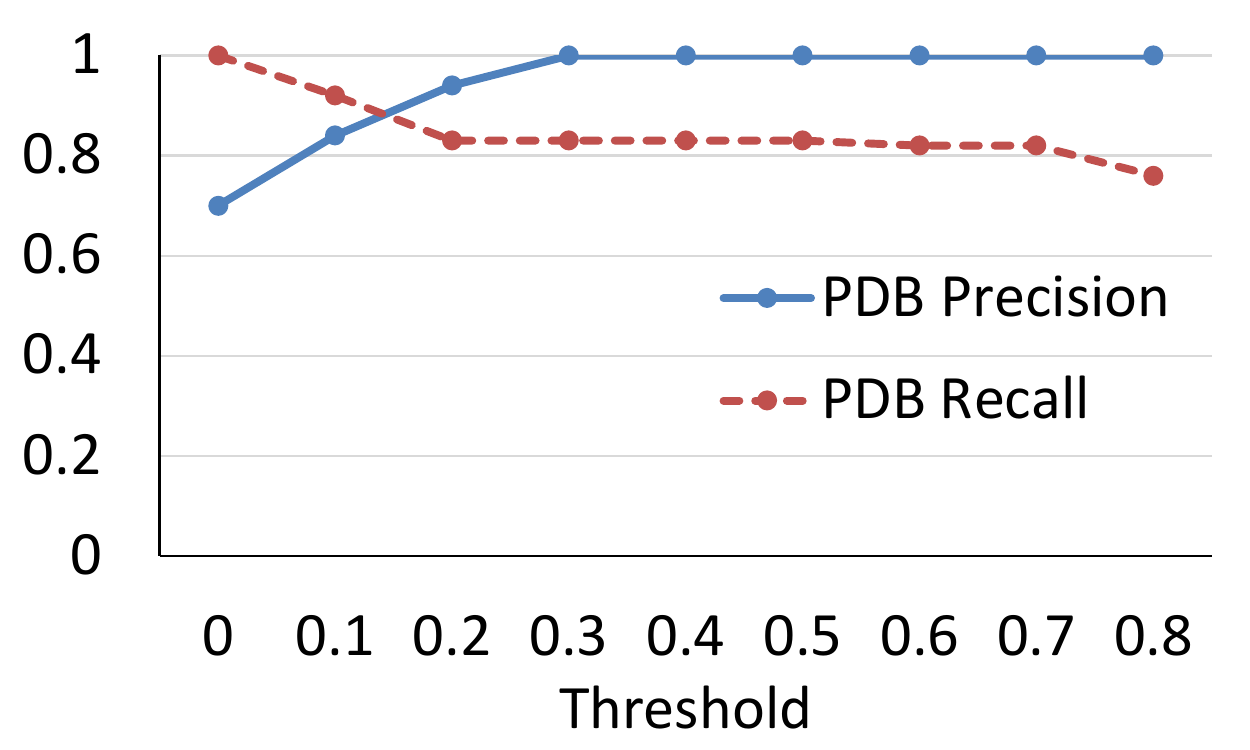}
   \caption{Precision and recall of the PDB method using a threshold.}
   \label{fig:pdb-threshold}
 \end{subfigure}
 ~
 \begin{subfigure}{0.20\textwidth}
   \centering
   \includegraphics[width=\textwidth]{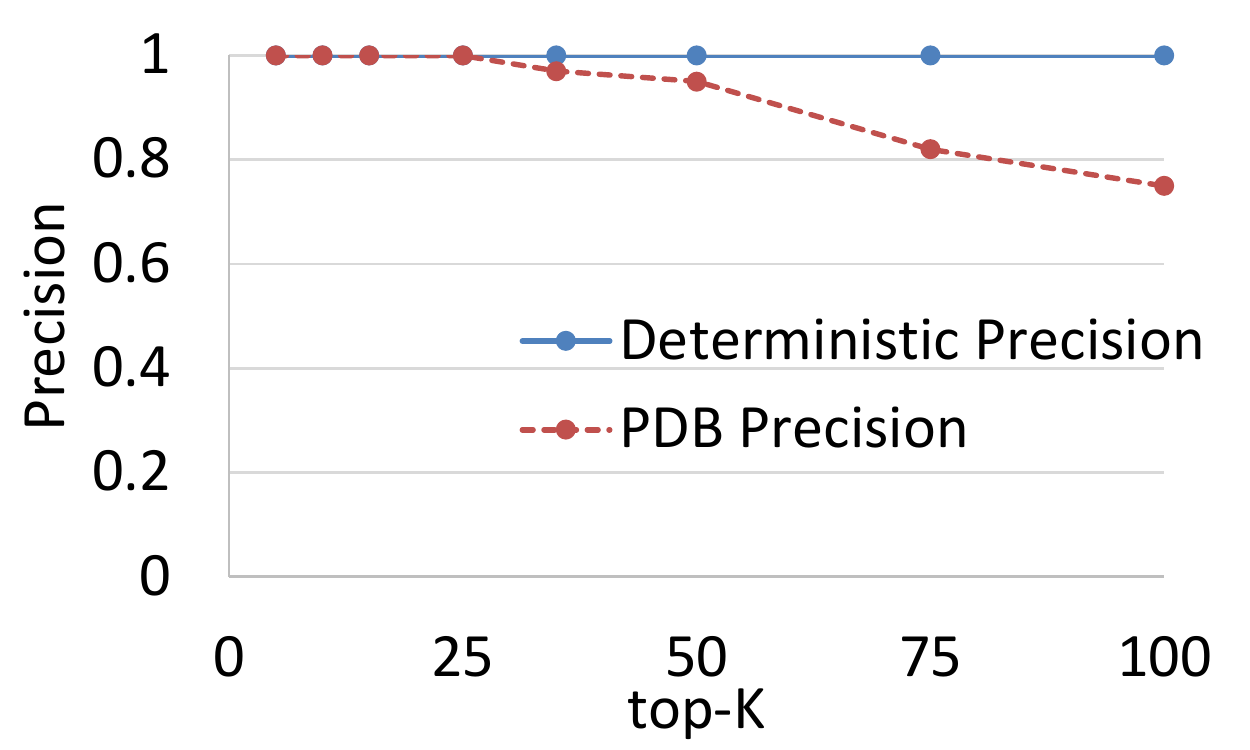}
   \caption{top-$k$ precision of PDB vs deterministic method.}
   \label{fig:pdb-topk}
 \end{subfigure}

 \caption{Results of probabilistic method.}
 \label{fig:pdb-expt}
 \end{figure*}
While Figure~\ref{fig-offline-clean-vs-cfd} showed only percentages, in
Figure~\ref{fig-offline-clean-vs-noise} we report the actual number of tuples cleaned in the
dataset along with the percentage cleaned. This curve shows that the raw number of tuples cleaned
always increases with higher input noise percentages.

\mund{Setting $\gamma$:} As explained in Section~\ref{sec:detdb},
the weight given to the edit distance ($\delta$) compared to the weight
given to the distributional similarity ($1 - \delta$); and the overcorrection parameter
($\gamma$) are parameters that can be tuned, and should be set
based on which kind of error is more likely to occur. In our experiments, we performed a grid search
to determine the best values of $\delta$ and $\gamma$ to use. In Figure~\ref{fig:gain-vs-parameter},
we show a portion of the grid search where $\delta = 2/5$, and varying $\gamma$.

\begin{figure*}[ht]
\centering
\begin{subfigure}{0.28\textwidth}
    \centering
    \includegraphics[width=\textwidth]{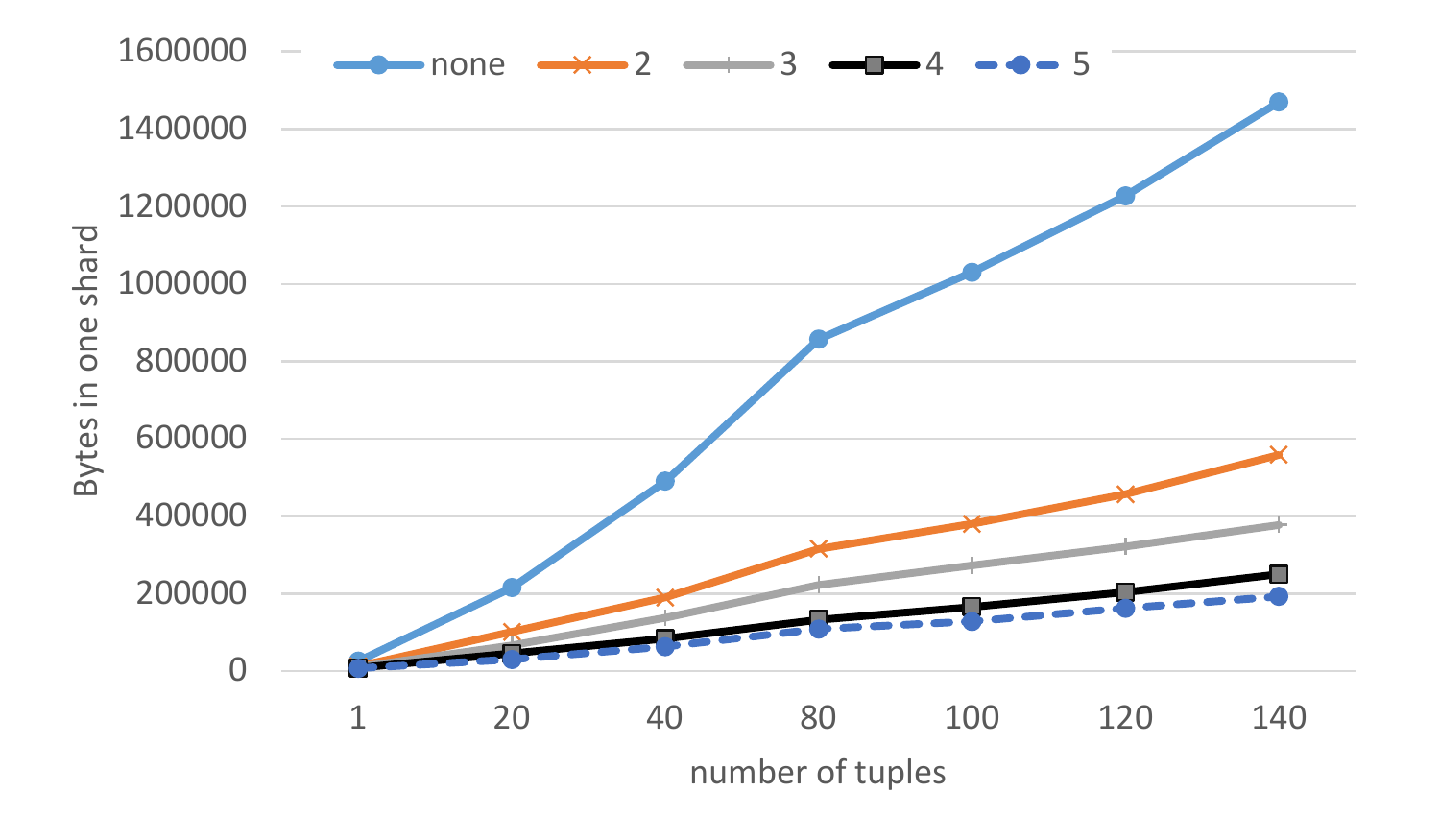}
    \caption{vs the Number of Tuples (in Thousands) in the Dataset, for Various Number of Shards.}
    \label{fig:shards-tuples}
\end{subfigure}~~
\begin{subfigure}{0.28\textwidth}
    \centering
    \includegraphics[width=\textwidth]{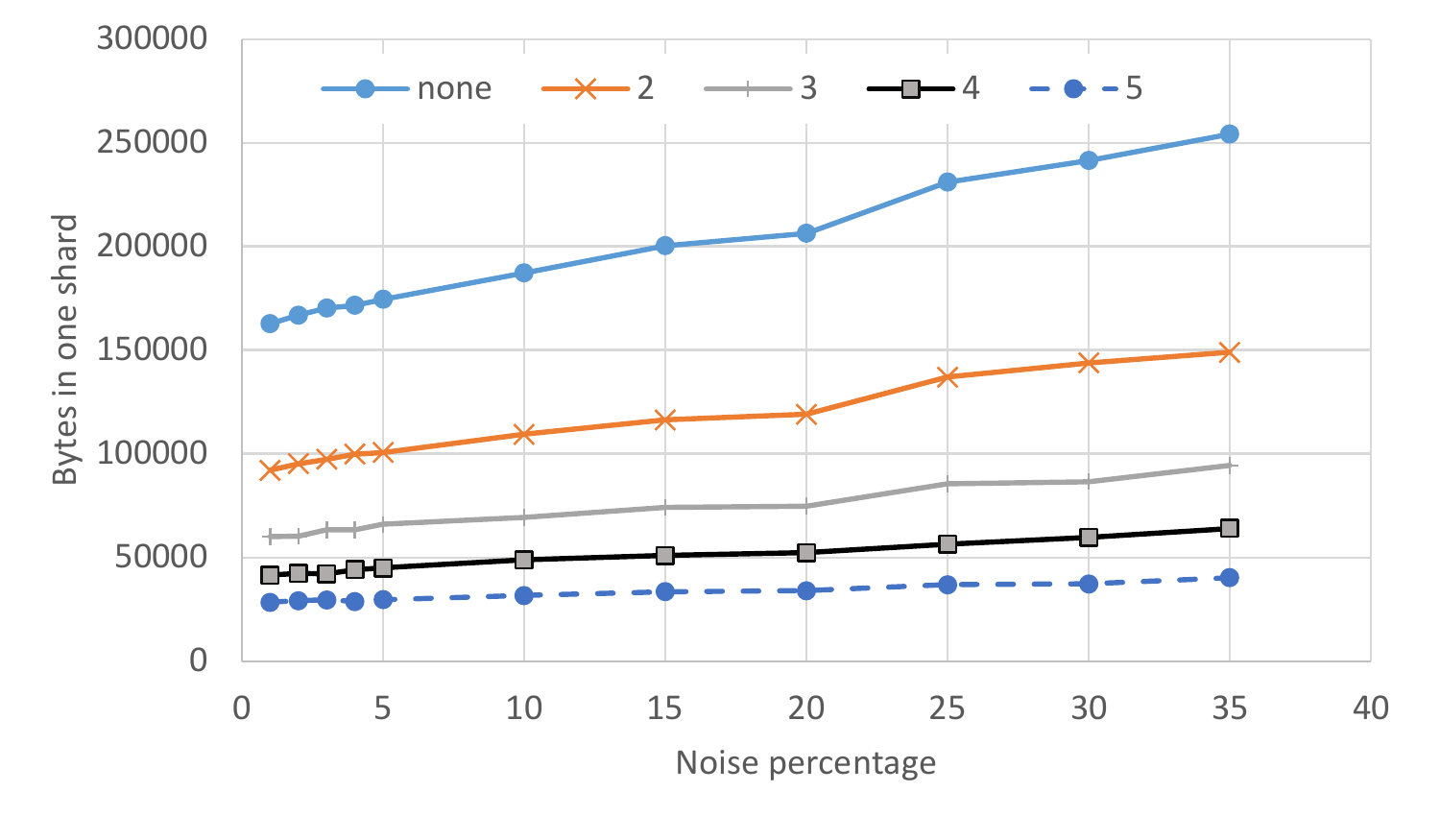}
    \caption{vs the Noise in the Dataset, for Various Number of Shards.}
    \label{fig:shards-noise}
\end{subfigure}~~
\caption{Map-Reduce index sizes}
\label{fig-mr-expts}
\begin{subfigure}{0.3\textwidth}
    \centering
    \includegraphics[width=\textwidth]{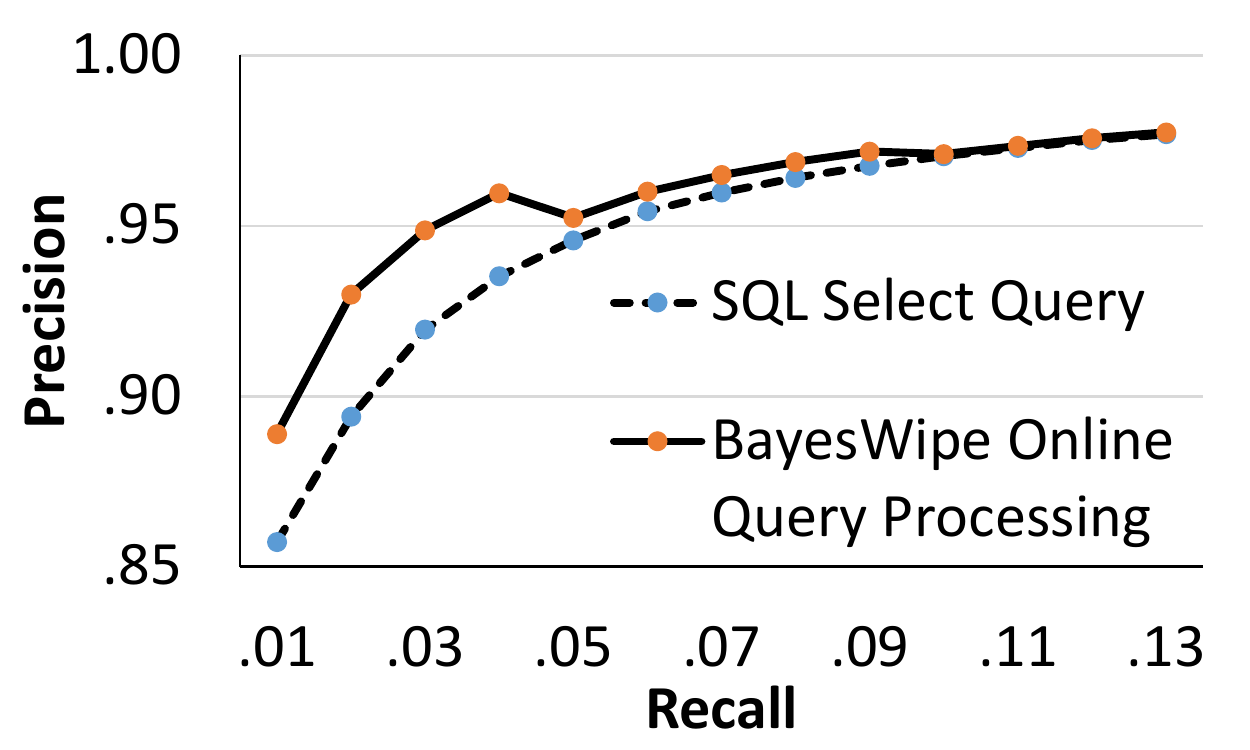}
    \caption{Average precision vs recall} 
    \label{fig:precision-20-noise}
\end{subfigure}~~
\begin{subfigure}{0.3\textwidth}
    \centering
    \includegraphics[width=\textwidth]{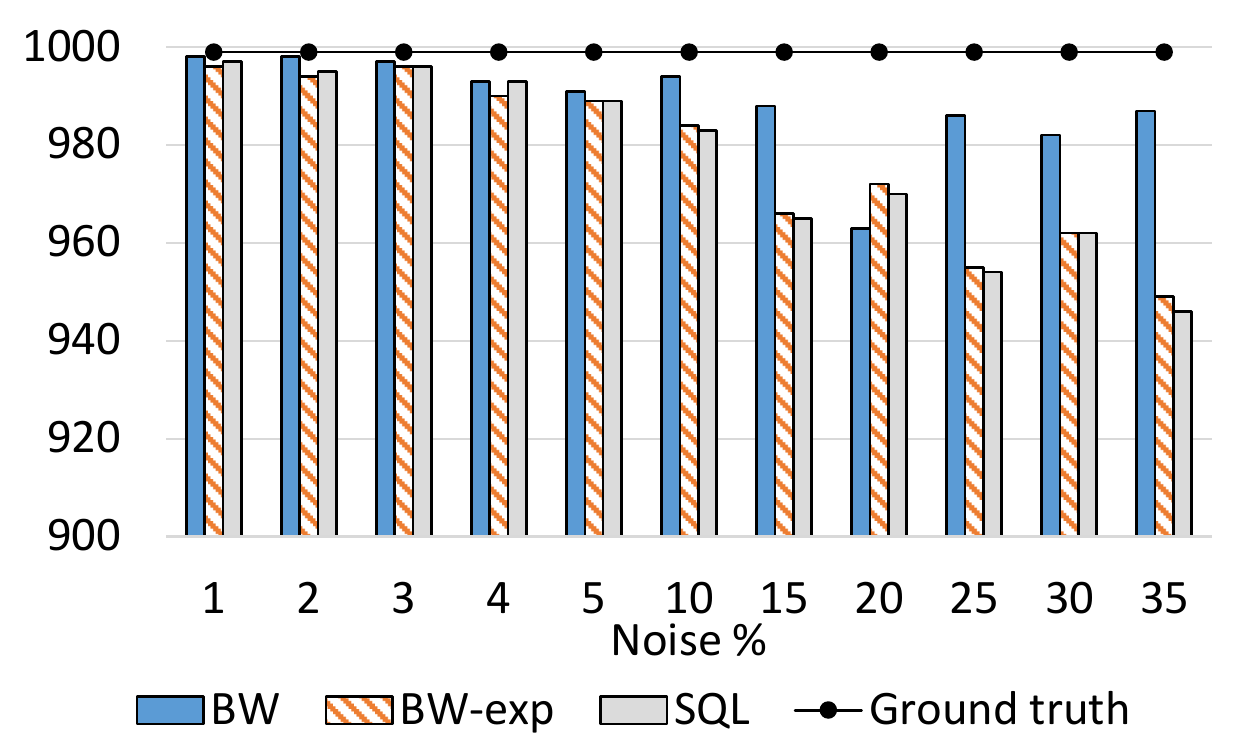}
    \caption{Net improvement in data quality (TP-FP)}
   \label{fig:online-ablution-net-improvement}
\end{subfigure}~~
\caption{Online cleaning  mode of \bw}
\label{fig-online-cleaning}
\vspace{-4mm}
\end{figure*}

\begin{figure*}[ht]
\centering
  \begin{subfigure}{0.22\textwidth}
    \centering
    \includegraphics[width=\textwidth]{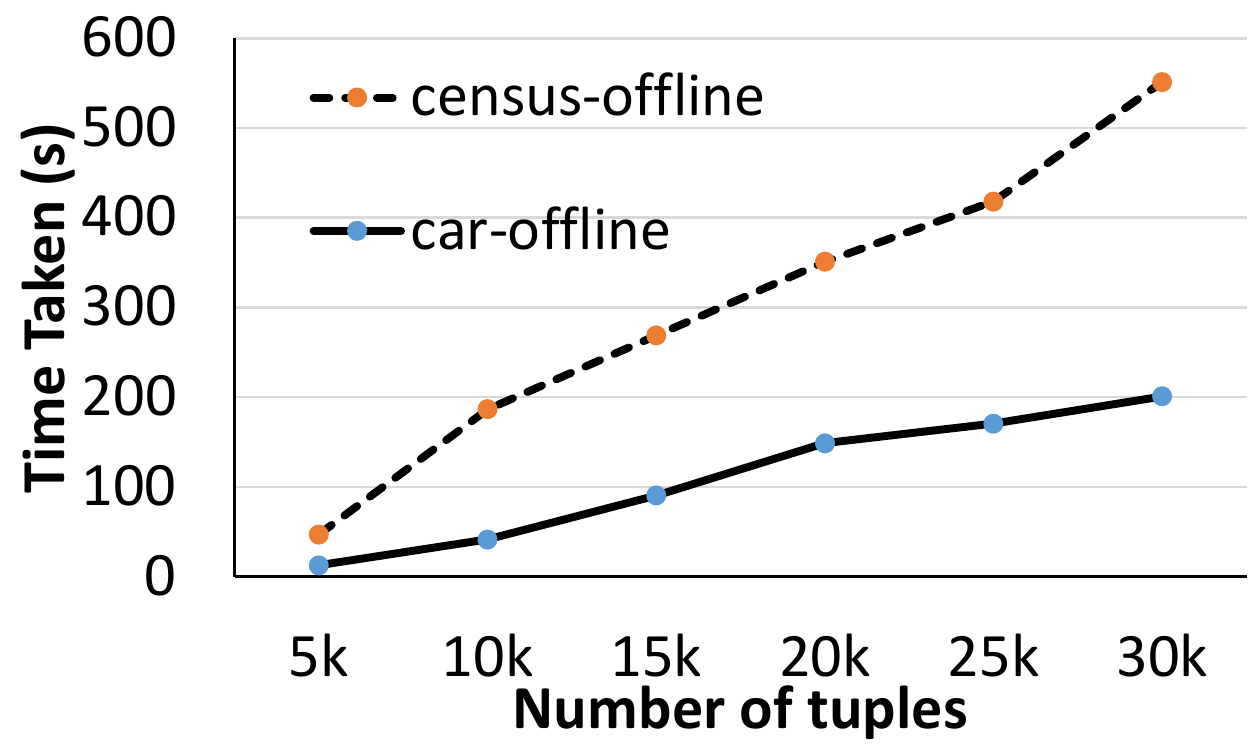}
    \caption{Time taken vs number of tuples in offline mode}
    \label{fig:offline-time-vs-tuples}
  \end{subfigure}~~
  \begin{subfigure}{0.22\textwidth}
    \centering
    \includegraphics[width=\textwidth]{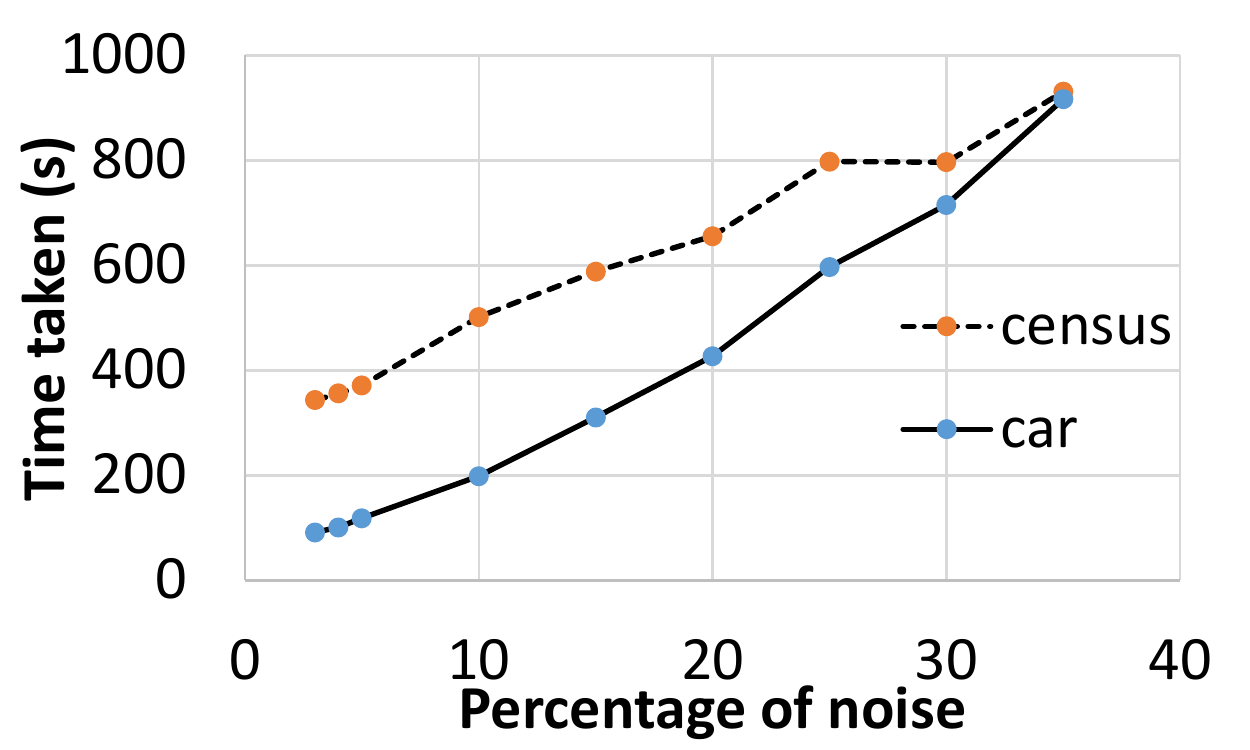}
    \caption{Time taken vs noise percentage in offline mode}
    \label{fig:offline-time-vs-noise}
  \end{subfigure}~~
  \begin{subfigure}{0.22\textwidth}
    \centering
    \includegraphics[width=\textwidth]{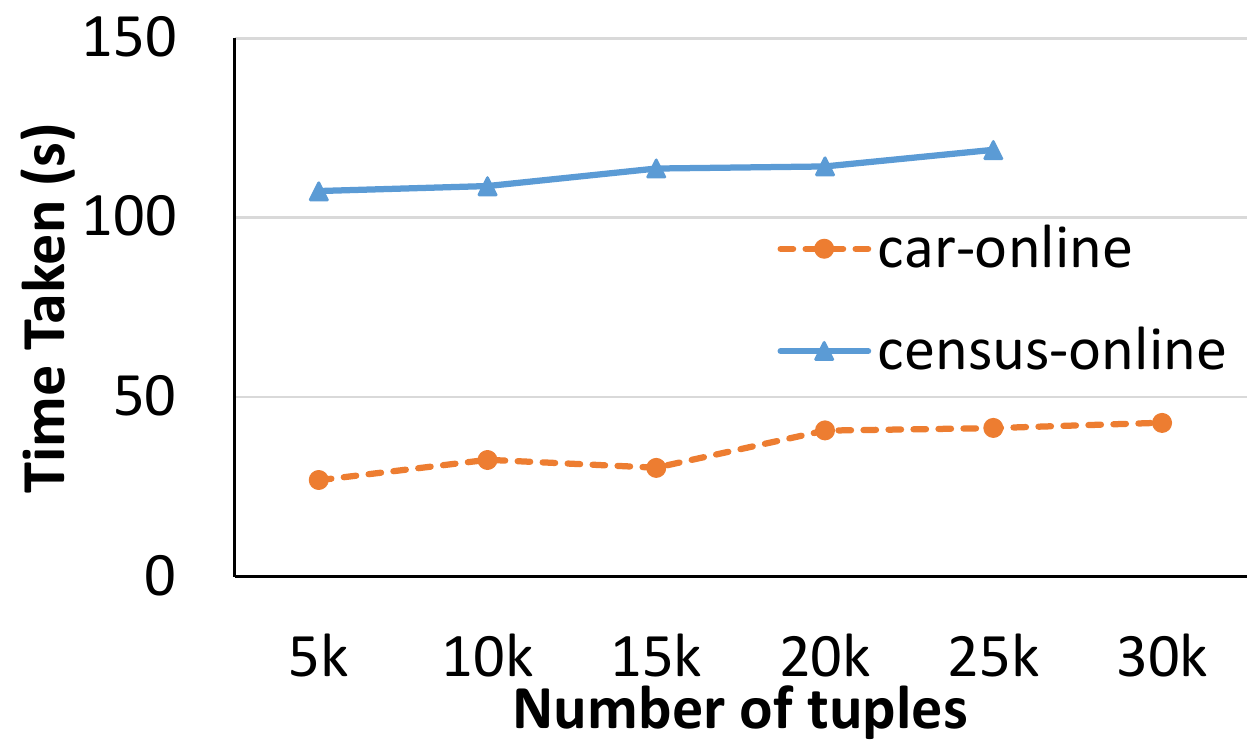}
    \caption{Time taken vs number of tuples in online mode}
    \label{fig:online-time-vs-tuples}
  \end{subfigure}~~
  \begin{subfigure}{0.22\textwidth}
    \centering
    \includegraphics[width=\textwidth]{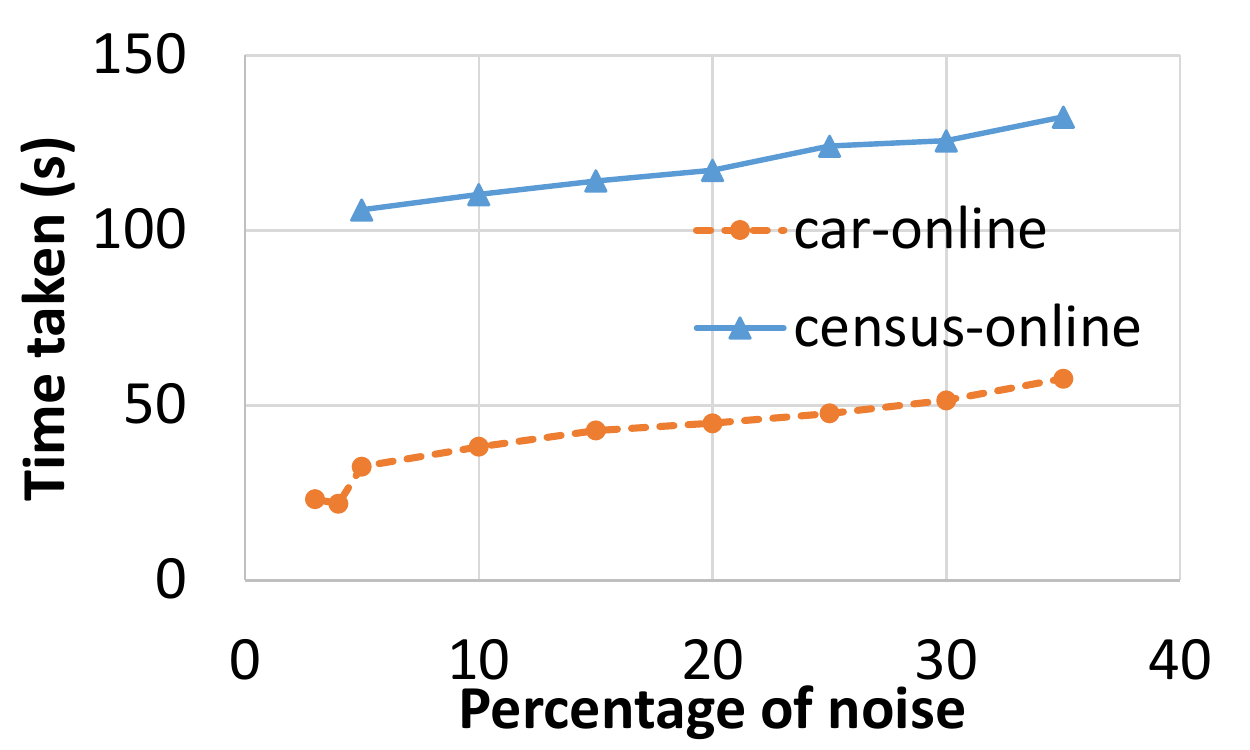}
    \caption{Time taken vs noise percentage in online mode}
    \label{fig:online-time-vs-noise}
  \end{subfigure}
  \caption{Performance evaluations}
  \vspace{-4mm}
  \label{fig:efficiency}
\end{figure*}

The ``values corrected'' data points in the graph correspond to the number of erroneous
attribute values that the algorithm successfully corrected (when checked
against the ground truth).  The ``false positives'' are the number of legitimate values that the
algorithm changes to an erroneous value. When cleaning the data, our
algorithm chooses a candidate tuple based on both the prior of the
candidate as well as the likelihood of the correction given the
evidence. Low values of $\gamma$ give a higher weight to
the prior than the likelihood, allowing tuples to be changed more easily
to candidates with high prior.
The ``overall gain'' in the number of clean values is calculated as the
difference of clean values between the output and input of the algorithm.

If we set the parameter values too low, we will correct most wrong tuples
in the input dataset, but we will also `overcorrect' a larger number of tuples.
If the parameters are set too high, then the system will not correct
many errors --- but the number of `overcorrections' will also be lower.
Based on these experiments, we picked a parameter value of
$\delta=0.638, \gamma=5.8$ and kept it constant throughout.

 \mund{Using probabilistic databases:} 
 We empirically evaluate the PDB-mode of \bw\ in Figure~\ref{fig:pdb-expt}. 
 In the first figure, we show the performance of the PDB mode of \bw\ against the deterministic
 mode for specific queries. As we can see from the first, third and seventh queries, the \bw-PDB
 improves the recall without any loss of precision. However, in most cases (and on average),
 \bw-PDB provides a better recall at the cost of some precision.

 The second figure shows the performance of \bw-PDB as the probability threshold for inclusion of a tuple
 in the resultset is varied. As expected, with low values of the threshold, the system allows 
 most tuples into the resultset, thus showing high recall and low precision. As the threshold
 increased, the precision increases, but the recall falls. 

 In Figure~\ref{fig:pdb-topk}, we compare the precision of the PDB mode using top-$k$ determinization
 against the deterministic mode of \bw. As expected, both the modes show 
 high precision for low values of $k$,
 indicating that the initial results are clean and relevant to the user. For higher values of $k$,
 the PDB precision falls off, indicating that PDB methods are more useful for scenarios where 
 high recall is important without sacrificing too much precision.

\mund{Online Query Processing:}
\label{sec:expt-qr}
While in the offline mode, we had the luxury of changing the tuples in the database itself, in online query processing, we use query rewriting to obtain a resultset that is similar to the offline results, without modification to the database.
We consider an SQL select query system as our baseline. We evaluate the precision
and recall of our method against the ground truth and compare it with the baseline, using
randomly generated queries.

We issued randomly generated queries to both \bw\ and the baseline system.
Figure~\ref{fig:precision-20-noise} shows the average precision over 10 queries
at various recall values. It shows that our system outperforms the SQL select query
system in top-$k$ precision, especially since our system considers the relevance of the results
when ranking them. On the other hand, the SQL search approach is oblivious to ranking
and returns all tuples
that satisfy the user query. Thus it may return irrelevant tuples early on,  leading to less precision.

Figure~\ref{fig:online-ablution-net-improvement} shows the improvement in the
absolute numbers of tuples returned by the \bw\ system. The graph shows the
number of true positive tuples returned (tuples that match the query results
from the ground truth) minus the number of false positives (tuples that are
returned but do not appear in the ground truth result set). We also plot the
number of true positive results from the ground truth, which is the
theoretical maximum that any algorithm can achieve. The graph shows that the
\bw\ system outperforms the SQL query system at nearly every level of
noise. Further, the graph also illustrates that --- compared to an SQL query
baseline --- \bw\ closes the gap to the maximum possible number of tuples to
a large extent. In addition to showing the performance of \bw\ against the SQL
query baseline, we also show the performance of \bw\ without the query
relaxation part (called BW-exp\footnote{BW-exp stands for \bw-expanded,
since the only query rewriting operation done is query expansion.}). We can
see that the full \bw\ system
outperforms the BW-exp system significantly, showing that query relaxation
plays an important role in bringing relevant tuples to the resultset,
especially for higher values of noise.

This shows that our proposed query ranking strategy indeed captures
the expected relevance of the to-be-retrieved tuples, and the query rewriting module is able
to generate the highly ranked queries.

\mund{Efficiency:}
In Figure~\ref{fig:efficiency} we show the data cleaning time taken by the system in its various modes. The first two graphs show the offline mode, and the second two show the online mode. As can be seen from the graphs, \bw\ performs reasonably well both in datasets of large size and datasets with large noise.

\smallskip\noindent
{\bf Evaluation on real data with naturally occurring errors:}
In this section we used a dataset of 1.2 million tuples crawled from the cars.com
website\footnote{http://www.cars.com} to check the performance of the system with real-world
data, where the corruptions were not synthetically introduced. Since this data is large,
and the noise is completely naturally occurring, we do not have ground truth for this data.
To evaluate this system, we conducted an experiment on Amazon Mechanical Turk.
First, we ran the offline mode of \bw\ on the entire database. We then picked only those tuples
that were changed during the cleaning, and then created an interface in mechanical turk
where only those tuples were shown to the user in random order. Due to resource constraints,
the experiment was run with the first 200 tuples that the system found to be unclean.
We inserted 3 known answers into the questionnaire, and removed any responses that failed
to annotate at least 2 out of the 3 answers correctly.

\begin{figure}[t]
  \centering
  \includegraphics[width=0.5\textwidth]{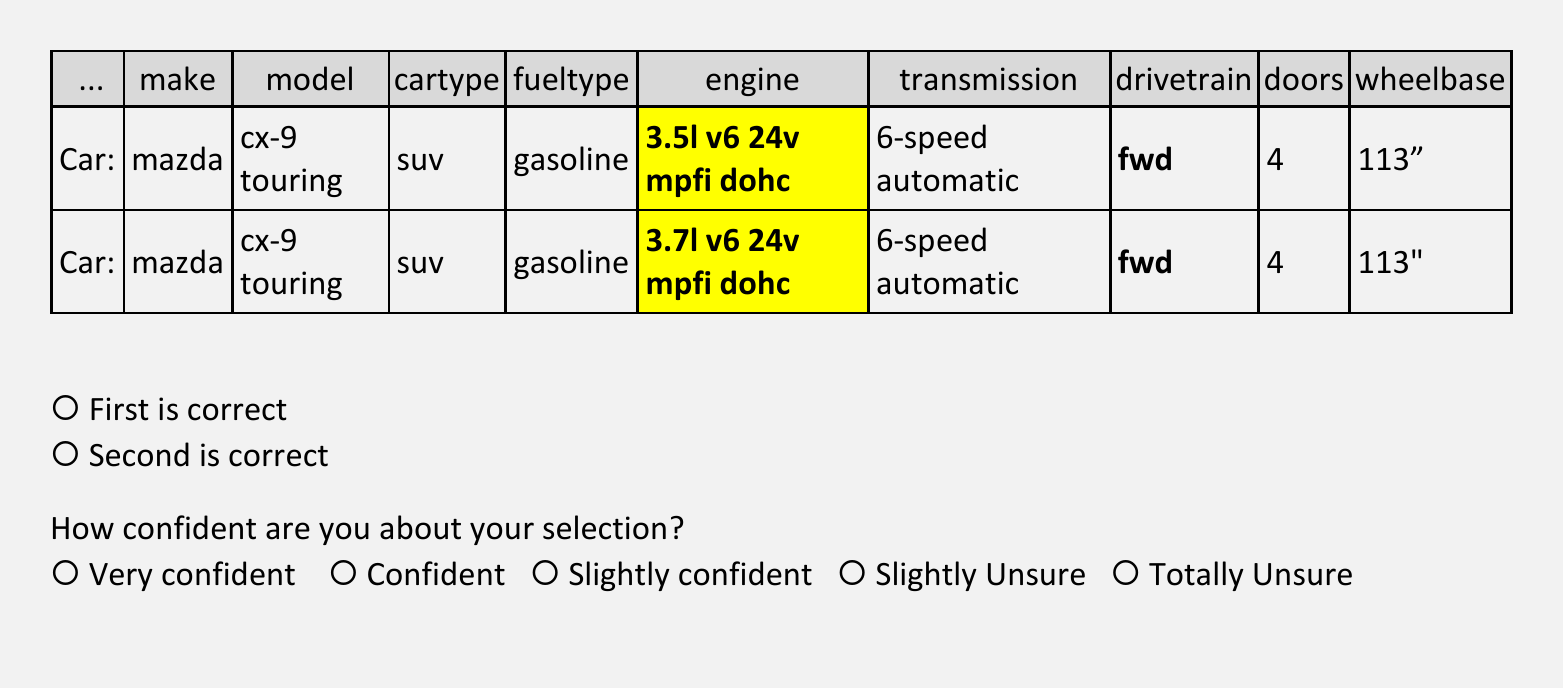}
  \vspace{-0.1in}
  \caption{A fragment of the questionnaire provided to the Mechanical
    Turk workers.}
  \mvp\mvp
  \label{fig:turk-questionnaire}
\end{figure}

An example is shown in Figure~\ref{fig:turk-questionnaire}. The turker is presented with
two cars, and she does not know which of the cars was originally present in the dirty dataset,
and which one was produced by \bw. The turker will use her own domain knowledge, or
perform a web search and discover that a ${\textsf{Mazda CX-9 touring}}$ is only available in
a ${\textsf{3.7l}}$ engine, not a ${\textsf{3.5l}}$.
Then the turker will be able to declare the second tuple as the correct option with
high confidence.

The results of this experiment are shown in Table~\ref{tbl:mturk-out}. As we can see, the users
consistently picked the tuples cleaned by \bw\ more favorably compared to the original dirty
tuples, proving that it is indeed effective in real-world datasets.
Notice that it is not trivial to obtain a 56\% rate of success in these experiments.
Finding a tuple which convinces the turkers that it is better than the original
requires searching through a huge space of possible corrections.
An algorithm that picks a possible correction randomly from this space is likely to get
a near 0\% accuracy.

The first row of Table~\ref{tbl:mturk-out} shows the fraction of tuples for which the turkers
picked the version cleaned by \bw\ and indicated that they were either `very confident' or `confident'.
The second row shows the fraction of tuples for all turker confidence values, and therefore is a
less reliable indicator of success.

In order to show the efficacy of \bw\ we also performed an
experiment in which the same tuples (the ones that \bw\ had changed) were
modified by a random perturbation. The random perturbation was done by the
same error process as described before (typo, deletion, substitution with
equal probability). Then these tuples (the original tuple from the database
and the perturbed tuple) were presented as two choices to the turkers. The
preference by the turkers for the randomly perturbed tuple over the original
dirty tuple is shown in the third column, `Random'. It is obvious from this
that the turkers overwhelmingly do not favor the random perturbed tuples. This
demonstrates two things. First, it shows the fact that \bw\ was performing
useful cleaning of the tuples. In fact, \bw\ shows a tenfold improvement over
the random perturbation model, as judged by human turkers. This shows that in
the large space of possible modifications of a wrong tuple, \bw\ picks the
correct one most of the time. Second, it provides additional support for the
fact that the turkers are picking the tuple carefully, and are not randomly
submitting their responses.

In this experiment, we also found the average fraction
of known answers that the turkers gave wrong answers to. This value was 8\%.
This leads to the conclusion that the difference between the turker's
preference of \bw\ over both the original tuples (which is 12\%) and the
random perturbation (which is 50\%) are both significant.

\begin{table}[t]
\centering
\begin{tabular}{|>{}m{0.6in} |>{}m{0.5in} | >{}m{0.4in}| >{}m{0.4in} |>{}m{0.7in} | } 
    \hline
    ${\textbf{Confidence}}$            & ${\textbf{\bw}}$ & ${\textbf{Original}}$ & ${\textbf{Random}}$ & ${\textbf{Increase over Random}}$\\ \hline
    High confidence only  & 56.3\% & 43.6\%   & 5.5\%    & 50.8\% points (10x better) \\ \hline
    All confidence values & 53.3\% & 46.7\%   & 12.4\% & 40.9\% points (4x better)  \\ \hline
  \end{tabular}
    \caption{Results of the Mechanical Turk Experiment, showing the percentage of tuples for which the users picked the results obtained by \bw\ as against the original tuple. Also shows performance against a random modification.}
    \label{tbl:mturk-out}
\end{table}

\section{Conclusion}
\label{sec:conclusion}
In this paper we presented a novel system, \bw\ that works using
end-to-end probabilistic semantics, and without access to clean master data.
We showed how to effectively learn the data source model as a Bayes network,
and how to model the error as a mixture of error features.
We showed the operation of this system in two modalities:
(1) offline data cleaning, an \emph{in situ} rectification of data and (2) online query
processing mode, a highly efficient way to obtain clean query results over inconsistent data.
 There is an option to generate a standard, deterministic database as the
output, as well as a probabilistic database, where all the alternatives are
preserved for further processing. 
We empirically showed that \bw\ outperformed existing baseline techniques in quality of results, and
was highly efficient. We also showed the performance of the \bw\ system at various stages of the query rewriting operation. We demonstrated how \bw\ can
be run on the map-reduce architecture so that it can scale to huge data sizes. User experiments showed that the system is useful in cleaning real-world
noisy data.
\vspace{-0.1in}


\vspace{-0.4in}
\begin{IEEEbiographynophoto}{Sushovan De}
is currently employed at Google. He graduated with a Ph.D. in Computer Science from Arizona State University. His research interests comprise 
Information Retrieval, Data Cleaning, and Probabilistic Databases. 
\end{IEEEbiographynophoto}
\vspace{-0.4in}
\begin{IEEEbiographynophoto}{Yuheng Hu}
is a Research Staff Member in the USER (User Systems and Experience Research) group at IBM Research - Almaden. He obtained his Ph.D in Computer Science at Arizona State University. 
His research interests are in the areas of Machine Learning, Social Computing and Human-Computer Interaction.
\end{IEEEbiographynophoto}
\vspace{-0.4in}
\begin{IEEEbiographynophoto}{Meduri Venkata Vamsikrishna}
is a Ph.D student in Computer Science at Arizona State University. He received a Master's degree in Computer Science from the National Unviersity of Singapore. His research interests 
include Data Mining from Social Media and Data Cleaning.
\end{IEEEbiographynophoto}
\vspace{-0.4in}
\begin{IEEEbiographynophoto}{Yi Chen}
 is an associate professor and the Henry J. Leir Chair in Healthcare in the School of Management with a joint appointment in the College of Computing Sciences at New Jersey Institute of Technology (NJIT).
She received her Ph.D. degree in Computer Science from the  University of Pennsylvania. 
Her current research focuses on Information Discovery on Big Data, Social Computing and Information Integration.
\end{IEEEbiographynophoto}
\vspace{-0.4in}
\begin{IEEEbiographynophoto}{Subbarao Kambhampati}
is a professor in Computer Science at Arizona State University. He directs the Yochan research group which is associated with the Artifical Intelligence Lab at Arizona State University. He is 
the "President-elect" of AAAI, the Association for the Advancement of Artificial Intelligence. He secured a Ph.D. degree in Computer Science from the University of Maryland, College Park.
His research interests are 
Automated Planning in Artificial Intelligence and  Data and Information Integration on the Web
\end{IEEEbiographynophoto}
\end{document}